\documentclass[traditabstract]{aa}  

\usepackage{graphicx}
\usepackage{txfonts}
\usepackage{longtable}
\usepackage{lscape}
\usepackage{makecell}
\usepackage{xcolor}
\usepackage{rotating}

\usepackage[breaklinks, colorlinks, citecolor=blue, urlcolor = blue]{hyperref}

\begin{document}

   \title{NOCTURNE. I. The radio spectrum of narrow-line Seyfert 1 galaxies}
   \titlerunning{NOCTURNE. I. The radio spectrum of NLS1s}
   \authorrunning{M. Berton et al.}
   \author{M. Berton
          \inst{1}\thanks{marco.berton@eso.org} \and
          E. Järvelä\inst{2} \and
          S. Chen\inst{3} \and
          L. Crepaldi\inst{4} \and
          I. Varglund\inst{5,6,7} \and
          M. Coloma Puga\inst{1,8,9} \and
          A. Jimenez-Gallardo\inst{1} \and
          A. L\"ahteenm\"aki \inst{5,6} \and
          S. Panda \inst{10} \and
          C. Piscitelli\inst{11,12} \and
          A. Tortosa \inst{13}
          }

   \institute{$^1$ European Southern Observatory (ESO), Alonso de Córdova 3107, Casilla 19, Santiago 19001, Chile \\
   $^2$ Department of Physics and Astronomy, Texas Tech University, Box 41051, Lubbock, 79409-1051, TX, USA; \\
   $^3$ Physics Department, Technion, Haifa 32000, Israel;\\
   $^4$ Dipartimento di Fisica e Astronomia ``G. Galilei", Universit\`a di Padova, Vicolo dell'Osservatorio 3, 35122, Padova, Italy; \\
   $^5$ Aalto University Metsähovi Radio Observatory, Metsähovintie 114, FI-02540 Kylmälä, Finland;\\
   $^6$ Aalto University Department of Electronics and Nanoengineering, P.O. Box 15500, FI-00076 AALTO, Finland;\\
   $^7$ Centre for Astrophysics Research, University of Hertfordshire, College Lane, Hatfield AL10 9AB, UK; \\
   $^{8}$ Dipartimento di Fisica, Universit\`a degli Studi di Torino, Via Pietro Giuria 1, 10125 (Torino), Italy; \\  
   $^{9}$ INAF - Osservatorio Astrofisico di Torino, Via Osservatorio 20, I-10025 Pino Torinese,
  Italy; \\
   $^{10}$ International Gemini Observatory/NSF NOIRLab, Casilla 603, La Serena, Chile; \\ 
   $^{11}$ Department of Physics, Astronomy Section, University of Trieste, Via G.B. Tiepolo, 11, I-34143 Trieste, Italy; \\
   $^{12}$ INAF - Osservatorio Astronomico di Trieste, Via G. B. Tiepolo 11, I-34143 Trieste, Italy; \\
   $^{13}$ INAF - Osservatorio Astronomico di Roma, Via Frascati 33, I-00040 Monte Porzio Catone, Italy. \\
}

   \date{Received \today; accepted }

\abstract
   {The origin of the radio emission in active galactic nuclei (AGN) is still debated. Multiple physical mechanisms can contribute to the spectrum at these frequencies, including relativistic jets, the jet base, outflows, star formation, and synchrotron emission from the hot corona. Recently, new extreme radio variability has been observed in the class of low-mass/high-Eddington AGN known as narrow-line Seyfert 1 (NLS1) galaxies, suggesting that another, more exotic mechanism may also play a role, especially at frequencies above 10~GHz. To investigate this relatively unexplored area of the radio spectrum, we observed a sample of 50 NLS1s with the Karl G. Jansky Very Large Array (JVLA), and 20 of them were observed twice. In this sample, 24 sources were not detected, while the others are typically characterized by a steep spectrum that can be modeled with a power law. We also identified two new candidate jetted NLS1s, including a high-frequency peaker, which is an extremely young relativistic jet. We found no significant variability in the sources observed twice. We conclude that the radio spectrum of NLS1s is typically dominated by optically thin emission, likely from low-power outflows, or by circumnuclear star formation, with a limited contribution from relativistic jets. Further studies at different spatial scales and at other wavelengths are necessary to fully constrain the origin of the radio emission in this class of active galaxies.}

\keywords{Active galactic nuclei -- Radiation mechanisms: non-thermal -- Galaxies: active -- Galaxies: jets -- Radio continuum: galaxies                }

   \maketitle
%
\newcommand{\kms}{km s$^{-1}$}
\newcommand{\ergs}{erg s$^{-1}$}
\newcommand{\chired}{$\chi^2_\nu$}
\newcommand{\hb}{H$\beta$}

\section{Introduction}
\nolinenumbers
In the last decades, the physical properties of active galactic nuclei (AGN) with faint radio emission have been the subject of a long-standing debate \citep{Panessa19}. The origin of this debate, however, lies in the ill-defined parameter called radio-loudness (RL, \citealp{Kellermann89}), which classifies AGN as radio-loud or radio-quiet. Initially, it was defined as the ratio between the radio flux density at 5~GHz and the optical B-band magnitude, but several alternative versions based on the same concept have been used in the literature \citep{Ganci19, Gloudemans21}. Usually, sources whose RL$>$10 are considered radio-loud and are associated with the presence of powerful relativistic jets; otherwise, they are radio-quiet, with unclear origin of the radio emission (e.g., weaker jets, star formation, coronal emission). While in some cases this parameter can provide a rough idea of the physical properties of AGN, multiple lines of observational evidence show that sources do not follow this simple bimodality \citep[e.g.,][]{Arsenov25}. Several years ago, it was already noted that this parameter is strongly dependent on the way it is measured, since a different aperture on resolved objects could cause a source to go from radio-quiet to radio-loud \citep{Ho01}. However, the strongest arguments against the use of radio loudness come from a particular class of AGN, that of narrow-line Seyfert 1 (NLS1) galaxies \citep{Berton21c}. 

NLS1s were classified by \citet{Osterbrock85}, and are identified based on the optical spectrum: the full-width at half maximum (FWHM) of their broad H$\beta$ emission line is $<$2000 km s$^{-1}$ \citep{Goodrich89}, and their [O~III] emission is weak compared to the broad H$\beta$ ($S$([O~III])/$S$(H$\beta$)$<$3). They often exhibit strong Fe~II emission, although this is not always the case \citep{Pogge11, Cracco16}. The narrow FWHM(H$\beta$) is typically interpreted as a sign of low rotational velocity around a low-mass black hole (10$^{6}$--10$^{8} M_{\odot}$, \citealp{Peterson11, Komossa18}), often leading to high Eddington ratios \citep{Boroson92, Sulentic00, Marziani01, Marziani18b, Marziani25, Panda19}. These properties, and the prevalence of spiral galaxies among their hosts \citep{Jarvela18, Berton19a, Olguiniglesias20, Varglund22, Vietri22, Varglund23}, have led to the conclusion that they are AGN in an early stage of their evolution \citep{Mathur00, Berton17, Fraixburnet17b}, possibly experiencing one of their first activity cycles and therefore the low-redshift analogs of type 1 AGN observed in the early Universe \citep{Maiolino25, Berton25}. Interestingly, often NLS1s do not obey the radio-loudness dichotomy. In some sources, the star formation is so strong that they can appear radio-loud even without harboring a relativistic jet \citep{Caccianiga15}. In other cases, their relativistic jets are rather faint compared to the optical, thus leading to a radio-quiet classification \citep{Vietri22, Wang25}. Finally, their radio emission can be dramatically variable, leading to a different classification based on the observation epoch and frequency \citep{Lahteenmaki18, Berton20b, Jarvela21, Jarvela24}. 

The discovery of these extreme flares at 37~GHz came from the Mets\"ahovi Radio Observatory (MRO), when a sample of NLS1s with previously absent or weak radio emission, selected either based on their spectral energy distribution (SED) or large-scale environment \citep{Jarvela17}, were detected at Jy-level flux densities regardless of their selection criteria \citep{Lahteenmaki18}. The interpretation of this phenomenon is difficult, since the timescale of the flares is unprecedentedly short \citep{Jarvela24}, and these objects do not seem to share any distinguishing property, but appear as regular NLS1s, perhaps with higher than usual Eddington ratio \citep{Romano23, Crepaldi25}. Currently, the only known sources of this kind have been identified by their flaring activity at 37~GHz and, more sparsely, at 15~GHz with the Owens Valley Radio Observatory (OVRO). However, given the relatively high detection limit of MRO and the low number of OVRO detections, it is possible that many remain hidden within the general NLS1 population. For this reason, we adopted a different approach by studying a sample of southern NLS1s, selected by \citet{Chen18}, at high radio frequencies (10-35~GHz) to investigate the properties of the radio emission in these objects with the Karl G. Jansky Very Large Array (JVLA). 

This work belongs to a larger framework of studies, called NOCTURNE\footnote{\url{www.ejarvela.space/nocturne/}}, which stands for Narrow-line Seyfert 1 galaxies Over Cosmic Time: Unification, Reclassification, Nature, and Evolution. NOCTURNE is a panchromatic collaboration that aims to exploit the entire electromagnetic spectrum to understand several aspects of NLS1s better. This study continues the exploration of the radio properties of NLS1s, which until a decade ago were almost completely unexplored \citep[with a few exceptions, e.g. see,][]{Moran00, Komossa06, Yuan08}. The radio detection fraction of NLS1s depends on frequency, with the number of detections increasing toward lower radio frequencies due to the increasing contribution from star formation. Approximately 8\% of NLS1s are detected at 1.4~GHz in the Faint Images of the Radio Sky at Twenty-Centimeters (FIRST) survey \citep{Varglund25}, with an rms of 1~mJy. Their emission is typically dominated by star formation, that appears as a steep-spectrum patchy emission and is often detected in nearby objects. At times, they also show a non-negligible AGN contribution coming, for example, from the jet base or the corona \citep{Berton18a, Chen20, Chen22, Jarvela22}. However, in a few cases relativistic jets are present, mostly observed at small angles (i.e., blazar-like, \citealp{Abdo09c, Foschini11, Foschini15, Angelakis15, Foschini15, Jose24, Shao25}) and showing superluminal motion \citep{Lister16, Lister18}, but in some cases also seen at large angles \citep{Richards15, Congiu17, Congiu20, Vietri22, Chen24, Umayal25}. 
{All of these studies, however, were mostly focused on the low-frequency part of the radio spectrum. The coverage above $>$10 GHz is instead still sparse, and mostly dedicated to the gamma-ray emitting NLS1s \citep{Angelakis15, Lister16, Shao25}. Our study includes sources without any preselection, allowing us to filling this gap in our knowledge, and to better constrain the spectrum shape in this spectral range.}

The paper is organized as follows. In Sect. 2, we present our sample, the observations, and describe the data analysis. In Sect.~3, we present our results, describing the radio morphology and spectral properties of our sources. In Sect.~4, we discuss our findings, and in Sect.~5 we draw our conclusions. Throughout the paper, we adopt a standard $\Lambda$CDM cosmology, with a Hubble constant H$_0 = 70$ km s$^{-1}$\ Mpc$^{-1}$, and $\Omega_\Lambda = 0.73$ \citep{Komatsu11}. 
\begin{table*}[]
    \centering
    \footnotesize
    \caption{The sample.}
    \begin{tabular}{l c c c c c}
Short name & 6dFGS Name & R.A. {(J2000)} & Dec. {(J2000)} & z & Scale \\
\hline\hline
J0000-0541 & 6dFGS gJ000040.3-054101 & 00 00 40.27 & -05 41 01.0 & 0.093 & {1.733} \\
J0015-1509 & 6dFGS gJ001521.6-150951 & 00 15 21.63 & -15 09 51.0 & 0.078 & {1.478} \\
J0021-2050 & 6dFGS gJ002121.6-205018 & 00 21 21.55 & -20 50 17.6 & 0.183 & {3.088} \\
J0022-1039 & 6dFGS gJ002249.2-103956 & 00 22 49.16 & -10 39 55.8 & 0.414 & {5.535} \\
J0030-2028 & 6dFGS gJ003000.5-202856 & 00 30 00.52 & -20 28 56.2 & 0.289 & {4.368} \\
J0043-1655 & 6dFGS gJ004324.9-165557 & 00 43 24.91 & -16 55 57.1 & 0.330 & {4.787} \\
J0200-0845 & 6dFGS gJ020039.1-084555 & 02 00 39.12 & -08 45 55.0 & 0.432 & {5.678} \\
J0203-1247 & 6dFGS gJ020349.0-124717 & 02 03 49.03 & -12 47 16.8 & 0.052 & {1.015} \\
J0212-0201 & 6dFGS gJ021201.5-020154 & 02 12 01.47 & -02 01 53.8 & 0.437 & {5.716} \\
J0212-0737 & 6dFGS gJ021218.2-073720 & 02 12 18.33 & -07 37 19.8 & 0.173 & {2.951} \\
J0213-0551 & 6dFGS gJ021355.0-055121 & 02 13 55.16 & -05 51 21.3 & 0.139 & {2.460} \\
J0230-0859 & 6dFGS gJ023005.5-085953 & 02 30 05.52 & -08 59 53.3 & 0.016 & {0.326} \\
J0239-1118 & 6dFGS gJ023956.2-111813 & 02 39 56.15 & -11 18 13.0 & 0.203 & {3.353} \\
J0400-2500 & 6dFGS gJ040024.4-250044 & 04 00 24.40 & -25 00 44.3 & 0.097 & {1.799} \\
J0413-0050 & 6dFGS gJ041307.1-005017 & 04 13 07.05 & -00 50 16.6 & 0.040 & {0.792} \\
J0420-0530 & 6dFGS gJ042021.7-053054 & 04 20 21.74 & -05 30 54.4 & 0.199 & {3.301} \\
J0422-1854 & 6dFGS gJ042256.6-185442 & 04 22 56.56 & -18 54 42.3 & 0.064 & {1.232} \\
J0435-1643 & 6dFGS gJ043526.5-164346 & 04 35 26.50 & -16 43 46.0 & 0.098 & {1.816} \\
J0436-1022 & 6dFGS gJ043622.3-102234 & 04 36 22.24 & -10 22 33.8 & 0.035 & {0.697} \\
J0447-0403 & 6dFGS gJ044739.0-040330 & 04 47 39.02 & -04 03 29.7 & 0.081 & {1.530} \\
J0447-0508 & 6dFGS gJ044720.7-050814 & 04 47 20.73 & -05 08 14.1 & 0.044 & {0.867} \\
J0452-2953 & 6dFGS gJ045230.1-295335 & 04 52 30.10 & -29 53 35.3 & 0.285 & {4.325} \\
J0455-1456 & 6dFGS gJ045557.5-145641 & 04 55 57.52 & -14 56 41.2 & 0.136 & {2.415} \\
J0549-2425 & 6dFGS gJ054914.9-242552 & 05 41 58.03 & -37 38 37.1 & 0.224 & {3.619} \\
J0622-2317 & 6dFGS gJ062233.5-231742 & 06 22 33.53 & -23 17 41.7 & 0.037 & {0.735} \\
J0820-1741 & 6dFGS gJ082003.1-174151 & 08 20 03.10 & -17 41 50.9 & 0.073 & {1.391} \\
J0842-0349 & 6dFGS gJ084219.1-034931 & 08 42 19.11 & -03 49 31.4 & 0.357 & {5.042} \\
J0845-0732 & 6dFGS gJ084510.2-073205 & 08 45 10.25 & -07 32 05.2 & 0.103 & {1.897} \\
J0846-1214 & 6dFGS gJ084628.7-121409 & 08 46 28.67 & -12 14 09.3 & 0.107 & {1.962} \\
J0849-2351 & 6dFGS gJ084951.7-235125 & 08 49 51.67 & -23 51 25.0 & 0.127 & {2.278} \\
J0850-0318 & 6dFGS gJ085028.0-031817 & 08 50 27.96 & -03 18 16.7 & 0.162 & {2.796} \\
J1014-0418 & 6dFGS gJ101420.7-041841 & 10 14 20.68 & -04 18 40.3 & 0.058 & {1.125} \\
J1015-1652 & 6dFGS gJ101503.2-165214 & 10 15 03.20 & -16 52 14.0 & 0.432 & {5.678} \\
J1032-1609 & 6dFGS gJ103214.1-161000 & 10 32 14.13 & -16 09 59.7 & 0.052 & {1.015} \\
J1032-2707 & 6dFGS gJ103257.0-270730 & 10 32 57.04 & -27 07 30.3 & 0.071 & {1.356} \\
J1044-1826 & 6dFGS gJ104448.7-182653 & 10 44 48.72 & -18 26 53.2 & 0.113 & {2.059} \\
J1057-0805 & 6dFGS gJ105719.5-080541 & 10 57 19.45 & -08 05 40.5 & 0.221 & {3.582} \\
J1147-2145 & 6dFGS gJ114738.9-214508 & 11 47 38.87 & -21 45 07.7 & 0.219 & {3.557} \\
J2021-2235 & 6dFGS gJ202104.4-223518 & 20 21 04.38 & -22 35 18.3 & 0.185 & {3.115} \\
J2115-1417 & 6dFGS gJ211524.9-141706 & 21 15 24.88 & -14 17 05.7 & 0.271 & {4.171} \\
J2136-0116 & 6dFGS gJ213632.0-011626 & 21 36 32.02 & -01 16 26.1 & 0.273 & {4.193} \\
J2137-1112 & 6dFGS gJ213748.0-111204 & 21 37 47.95 & -11 12 03.6 & 0.113 & {2.059} \\
J2143-2958 & 6dFGS gJ214306.1-295817 & 21 43 06.10 & -29 58 16.0 & 0.120 & {2.169} \\
J2155-1210 & 6dFGS gJ215526.7-121032 & 21 55 26.74 & -12 10 31.7 & 0.086 & {1.615} \\
J2207-2824 & 6dFGS gJ220755.6-282406 & 22 07 55.62 & -28 24 06.2 & 0.178 & {3.020} \\
J2229-1401 & 6dFGS gJ222903.5-140106 & 22 29 03.51 & -14 01 06.2 & 0.236 & {3.766} \\
J2244-1822 & 6dFGS gJ224458.2-182250 & 22 44 58.19 & -18 22 49.5 & 0.198 & {3.288} \\
J2250-1152 & 6dFGS gJ225014.1-115201 & 22 50 14.06 & -11 52 00.8 & 0.118 & {2.138} \\
J2311-2022 & 6dFGS gJ231103.4-202221 & 23 11 03.36 & -20 22 20.6 & 0.121 & {2.185} \\
J2358-1028 & 6dFGS gJ235808.5-102843 & 23 58 08.46 & -10 28 43.1 & 0.167 & {2.867} \\
 \hline
    \end{tabular}
    \tablefoot{Columns: (1) Short name of the source, used throughout the paper; (2) Name of the source in the 6dFGS catalog; (3) right ascension; (4) declination; (5) redshift; (6) spatial scale (kpc/").}
    \label{tab:sample}
\end{table*}

\section{Observations and data analysis}

Currently, all of the known sources showing extreme flares are NLS1s. Since we have no way to know in advance which sources have the best chance of being detected, our strategy was to blindly observe NLS1s to study their spectral index above 15 GHz, i.e. OVRO already proved the detectability of some sources. In principle, the flaring NLS1s could show an inverted or flat spectrum in this spectral region. All the sources were derived from the sample selected by \citet{Chen18}. These sources, selected based on their optical spectra obtained by the 6 degree-field galaxy survey (6dFGS) are all in the Southern hemisphere, and have z$<$0.45. We selected only those sources with -30$^\circ <$ dec $<$ 0$^\circ$. Such sky positioning was chosen to maximize the chances for follow-up observations at all wavelengths, which is currently very unfavorable since all of the known objects have rather high declination ($>34^\circ$). Furthermore, some of these sources were already observed at 5~GHz with the JVLA in C configuration by \citet{Chen20}. Of the original sample of 192 objects, 87 meet our declination criterion. Out of those, 50 were observed with the JVLA, and 20 of them were observed twice, to maximize the chances of finding large-amplitude variability. The full sample is reported in Table~\ref{tab:sample}.


The sources were observed in C configuration, which was also used by \citet{Chen20}. We observed them in three different bands, Ku, K, Ka, centered at 15, 22, and 33 GHz, respectively (project VLA/22B-034, PI Berton). The observation dates for each source are reported in the Table~\ref{tab:data}. The nominal on-source time in each band was 2 minutes. The total bandwidth was 6~GHz in Ku and 8~GHz in K and Ka, each band divided into 128~MHz subbands, each consisting of 64 2~MHz channels. Each interval of right ascension was calibrated in flux and bandpass with a suitable source, and we used a different bright source as the complex gain calibrator for each one of the targets. The expected thermal noise were $\sim$15, 28, and 30 $\mu$Jy beam$^{-1}$ {at 15, 22, and 33 GHz,} respectively. These levels were reached or surpassed in most cases. We used the pre-processed science-ready data products (SRDP) provided in the National Radio Astronomy Observatory (NRAO) archive. The data were calibrated using the VLA Imaging Pipeline 2023.1.0.124, and we processed them using the Common Astronomy Software Applications (CASA) version 6.5.2.26. The data were manually checked by NRAO to produce the SRDP measurement set. We did not perform any additional flagging. We split the data for our sources from each measurement set, averaging over time (10~s) and frequency, to average 64 channels to a single output channel per spectral window. To create the radio maps, we used the CLEAN algorithm. To properly sample the beam, we adopted a cell size of 0.15", 0.10", and 0.07" {at 15, 22, and 33 GHz,} respectively. In no case was self-calibration necessary. All images are limited by noise, not by dynamic range. The final maps are not included in the paper, but they are available online. The measurements were performed by fitting the source with a two-dimensional Gaussian function. A single Gaussian was enough in all cases but three (see Sect.~\ref{sec:radio_morph}).
\begin{figure}[!t]
    \centering
    \includegraphics[width=\columnwidth, trim={0 11cm 0 1cm},clip]{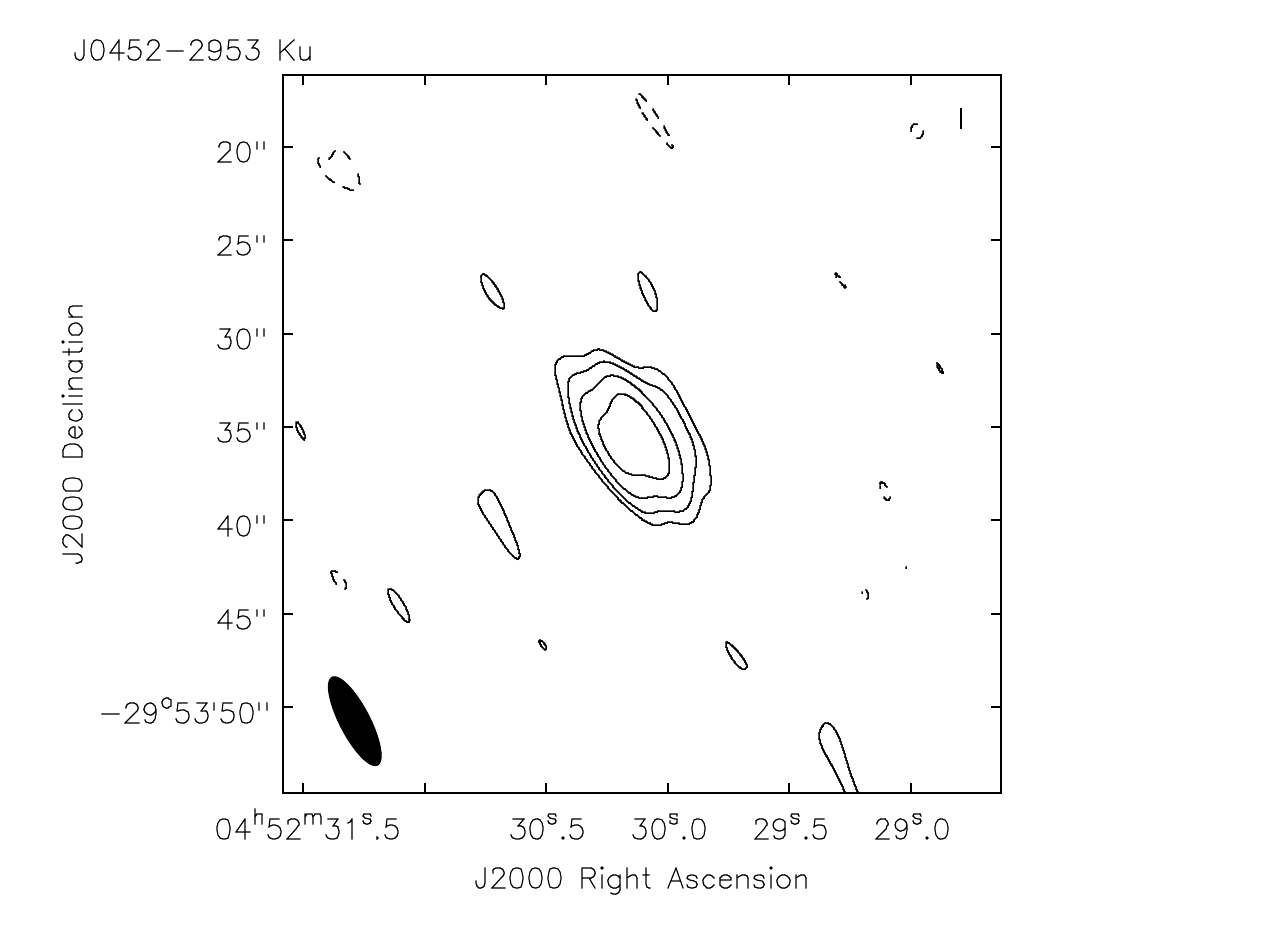}
    \caption{Radio map of J0452-2953 {at 15 GHz}. The map rms is $\sigma = 17 \mu$Jy, the contours are at [-3, 3, 6, 12, 24]$\times\sigma$.}
    \label{fig:J0452_map}
\end{figure}
\begin{figure}[!t]
    \centering
    \includegraphics[width=\columnwidth, trim={0 11cm 0 1cm},clip]{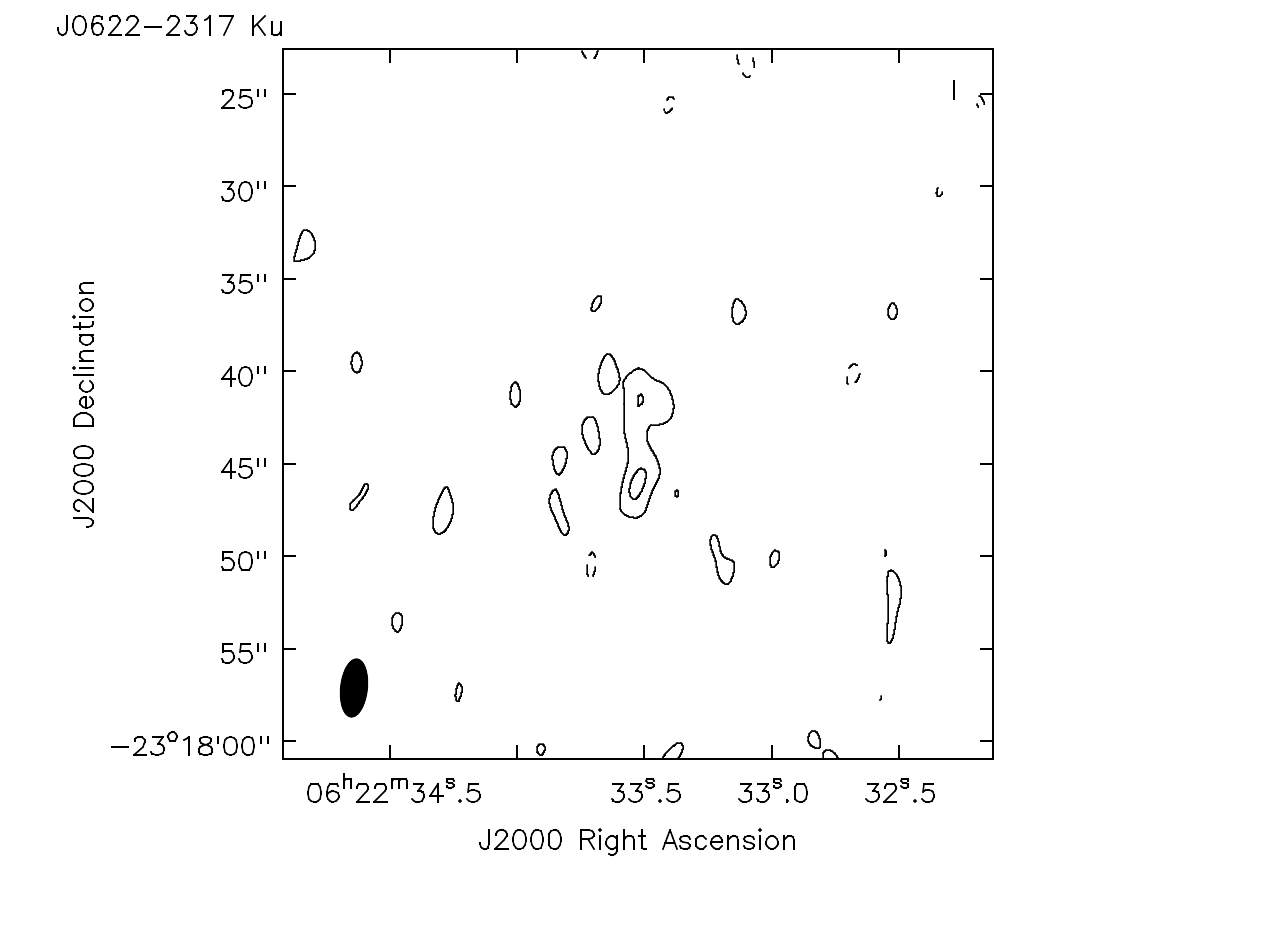}
    \caption{Radio map of J0622-2317 {at 15 GHz}. The map rms is $\sigma = 12 \mu$Jy, the contours are at [-3, 3, 6]$\times\sigma$.}
    \label{fig:J0622_map}
\end{figure}
\begin{figure}[!t]
    \centering
    \includegraphics[width=\columnwidth, trim={0 12cm 0 1cm},clip]{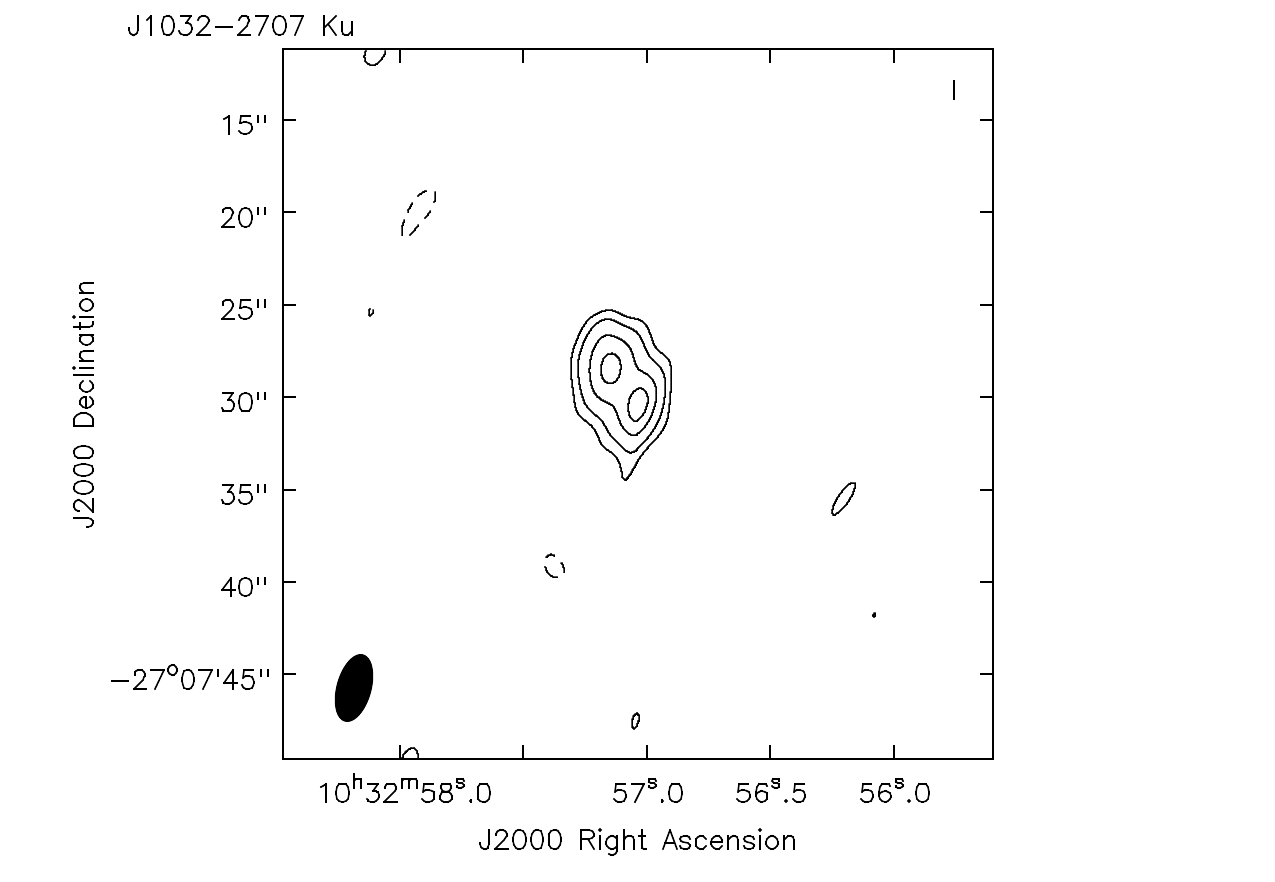}
    \caption{Radio map of J1032-2707 {at 15 GHz}. The map rms is $\sigma = 22 \mu$Jy, the contours are at [-3, 3, 6, 12, 18]$\times\sigma$.}
    \label{fig:J1032_map}
\end{figure}
We considered a source detected only when it showed at least two contours in our maps, that is, when the flux density is $>6\times$rms. Out of the 50 sources we observed, 24 were not detected at any frequency. We report the upper limit on their flux density as $6\times$rms in Table~\ref{tab:nodetection}. The remaining 26 sources, instead, were detected at least in one band. Their flux densities, derived from fitting with a single Gaussian, are reported in Table~\ref{tab:data}, while their luminosities are shown in Table~\ref{tab:luminosity}. To fully characterize their radio spectrum, we retrieved archival data from various surveys, that is the FIRST survey \citep{Becker95}, the National Radio Astronomy Observatory (NRAO) Very Large Array (VLA) Sky Survey (NVSS,  \citealp{Condon98}), the Very Large Array Sky Survey (VLASS, \citealp{Lacy20}), the Rapid Australian Square Kilometre Array Pathfinder (ASKAP) Continuum Survey (RACS, \citealp{Mcconnell20}), the Tata Institute of Fundamental Research (TIFR) Giant Metrewave Radio Telescope (GMRT) Sky Survey (TGSS, \citealp{Intema17}), the Australia Telescope 20 GHz Survey (AT20G, \citealp{Murphy10}), and finally from LOw Frequency ARray (LOFAR) Two-metre Sky Survey (LoTSS) \citep{Shimwell17}. For VLASS, we used both Epoch 1 and Epoch 2 data. Following \citet{Varglund25}, to find our sources, we ran a coordinate comparison with a tolerance radius of 15". All sources within this window we further checked to study whether or not there were multiple sources within the search region. For VLASS, there were some sources that were detected twice in the same epoch; for these, we have kept the source with the quality flag two, as this is the preferred detection based on the user guide\footnote{https://cirada.ca/vlasscatalogueql0}. All the additional archival measurements are reported in Table~\ref{tab:other_data}. 

For each source with at least one new JVLA detection, we used all the available data to model the spectrum with a power law and a log parabola, that is a line and parabola in the log-log plane. From a physical point of view, these fits allow us to test the presence of some curvature in the spectra, which may indicate synchrotron self-absorption (SSA) taking place at low frequencies. The fit with a power law provides a spectral index, defined as S$_\nu$ $\propto$ $\nu^{\alpha_\nu}$. For the parabola, defined in the log-log plane as \textit{f}($\log\nu$) = \textit{a}$\log\nu^2$ + \textit{b}$\log\nu$ + \textit{c}, we forced the \textit{a} coefficient to be negative, otherwise the model would diverge at low frequency. In some cases, the \textit{a} coefficient is consistent with zero, showing that the linear model is already sufficient to reproduce the data. When the two models produced significantly different results, we performed an F test to compare them and determine whether adding one order to the fitting function improves the fit. The results are reported in Table~\ref{tab:spind}.

\begin{table*}[]
    \centering
    \footnotesize
    \caption{Measurements for all the sources detected at least at one frequency in one epoch.}
    \begin{tabular}{l c c c c c c c c c c}
Name & MJD & rms (Ku) & S$_p$ (Ku) & S$_i$ (Ku) & rms (K) & S$_p$ (K) & S$_i$ (K) & rms (Ka) & S$_p$ (Ka) & S$_i$ (Ka) \\
\hline\hline
J0000-0541 & 59904 & 15 & 377$\pm$24 & 568$\pm$46 & 18 & 207$\pm$16 & 492$\pm$48 & 25 & 172$\pm$16 & 158$\pm$24\\
J0022-1039 & 59904 & 15 & 170$\pm$19 & 278$\pm$47 & 17 & 108$\pm$10 & 129$\pm$20 & 25 & $<$150 & {} \\
J0203-1247 & 59863 & 17 & 250$\pm$20 & 276$\pm$33 & 27 & 141$\pm$15 & 156$\pm$27 & 35 & 227$\pm$23 & 323$\pm$51\\
J0203-1247 & 59902 & 13 & 209$\pm$15 & 294$\pm$29 & 19 & 182$\pm$15 & 203$\pm$26 & 25 & 170$\pm$12 & 185$\pm$20\\
J0212-0201 & 59863 & 18 & 111$\pm$8 & 79$\pm$11 & 31 & $<$186 & {} & 40 & $<$240 & {} \\
J0212-0201 & 59902 & 13 & 98$\pm$9 & 94$\pm$15 & 22 & $<$132 & {} & 25 & $<$150 & {} \\
J0213-0551 & 59863 & 15 & 95$\pm$11 & 95$\pm$19 & 30 & $<$180 & {} & 35 & $<$210 & {} \\
J0213-0551 & 59902 & 15 & 104$\pm$11 & 130$\pm$21 & 22 & $<$132 & {} & 25 & $<$150 & {} \\
J0230-0859 & 59863 & 19 & 512$\pm$31 & 593$\pm$46 & 30 & 409$\pm$30 & 598$\pm$60 & 39 & 261$\pm$29 & 382$\pm$61\\
J0230-0859 & 59902 & 15 & 473$\pm$26 & 552$\pm$36 & 22 & 290$\pm$22 & 450$\pm$45 & 25 & 251$\pm$21 & 290$\pm$37\\
J0239-1118 & 59863 & 17 & 405$\pm$25 & 456$\pm$37 & 27 & 550$\pm$43 & 555$\pm$67 & 39 & 667$\pm$46 & 719$\pm$71\\
J0239-1118 & 59902 & 15 & 398$\pm$23 & 368$\pm$27 & 25 & 443$\pm$37 & 547$\pm$68 & 27 & 578$\pm$42 & 620$\pm$67\\
J0400-2500 & 59872 & 20 & 416$\pm$25 & 501$\pm$41 & 35 & 254$\pm$22 & 390$\pm$52 & 40 & $<$240 & {} \\
J0400-2500 & 59882 & 15 & 350$\pm$22 & 465$\pm$39 & 30 & 188$\pm$19 & 300$\pm$44 & 35 & $<$210 & {} \\
J0413-0050 & 59872 & 13 & $<$78 & {} & 19 & $<$114 & {} & 25 & $<$150 & {} \\
J0413-0050 & 59882 & 15 & 89$\pm$12 & 111$\pm$25 & * & * & * & 60 & $<$360 & {} \\
J0422-1854 & 59872 & 14 & 359$\pm$21 & 449$\pm$34 & 19 & 251$\pm$25 & 356$\pm$52 & 30 & 210$\pm$20 & 183$\pm$33\\
J0422-1854 & 59882 & 14 & 374$\pm$23 & 440$\pm$38 & 25 & 287$\pm$24 & 304$\pm$43 & 31 & $<$186 & {} \\
J0436-1022 & 59872 & 15 & 1707$\pm$87 & 1883$\pm$102 & 21 & 1177$\pm$61 & 1344$\pm$76 & 34 & 883$\pm$56 & 1050$\pm$89\\
J0436-1022 & 59882 & 18 & 1691$\pm$86 & 1830$\pm$97 & 27 & 1250$\pm$70 & 1351$\pm$92 & 36 & 1031$\pm$60 & 1178$\pm$84\\
J0447-0508 & 59872 & 27 & 1291$\pm$71 & 1612$\pm$101 & 20 & 822$\pm$48 & 1104$\pm$77 & 32 & 565$\pm$47 & 808$\pm$94\\
J0447-0508 & 59882 & 22 & 1405$\pm$75 & 1604$\pm$96 & 26 & 963$\pm$58 & 1237$\pm$94 & 36 & 572$\pm$45 & 672$\pm$79\\
J0452-2953 & 59872 & 17 & 722$\pm$43 & 1456$\pm$103 & 30 & 305$\pm$55 & 843$\pm$77 & 39 & 240$\pm$49 & 461$\pm$78\\
J0452-2953 & 59882 & 17 & 656$\pm$40 & 1296$\pm$91 & 26 & 394$\pm$69 & 757$\pm$188 & 36 & 277$\pm$32 & 384$\pm$77\\
J0549-2425 & 59898 & 15 & 593$\pm$31 & 648$\pm$38 & 20 & 450$\pm$27 & 491$\pm$39 & 30 & 366$\pm$27 & 382$\pm$41\\
J0549-2425 & 59899 & 15 & 593$\pm$32 & 685$\pm$44 & 20 & 414$\pm$30 & 492$\pm$51 & 30 & 332$\pm$24 & 400$\pm$43\\
J0622-2317 & 59898 & 12 & 83$\pm$12 & 189$\pm$26 & 17 & $<$102 & {} & 30 & $<$180 & {} \\
J0622-2317 & 59899 & 15 & $<$90 & {} & 17 & $<$102 & {} & 30 & $<$180 & {} \\
J0820-1741 & 59883 & 15 & 126$\pm$15 & 213$\pm$37 & 22 & $<$132 & {} & 33 & $<$198 & {} \\
J0820-1741 & 59904 & 17 & 169$\pm$14 & 219$\pm$29 & 22 & 139$\pm$17 & 202$\pm$40 & 32 & $<$192 & {} \\
J0842-0349 & 59883 & 17 & 117$\pm$14 & 252$\pm$40 & 30 & $<$180 & {} & 30 & $<$180 & {} \\
J0842-0349 & 59904 & 15 & 108$\pm$14 & 186$\pm$36 & 25 & $<$150 & {} & 32 & $<$192 & {} \\
J0846-1214 & 59883 & 21 & 2051$\pm$111 & 2090$\pm$132 & 27 & 1381$\pm$78 & 1486$\pm$103 & 30 & 785$\pm$51 & 1039$\pm$89\\
J0846-1214 & 59904 & 15 & 2099$\pm$112 & 2148$\pm$109 & 22 & 1402$\pm$74 & 1499$\pm$89 & 30 & 899$\pm$51 & 992$\pm$71\\
J0849-2351 & 59883 & 16 & 343$\pm$21 & 414$\pm$34 & 25 & 165$\pm$23 & 273$\pm$57 & 35 & $<$210 & {} \\
J0849-2351 & 59904 & 16 & 354$\pm$21 & 457$\pm$37 & 21 & 191$\pm$17 & 257$\pm$37 & 31 & 200$\pm$21 & 307$\pm$50\\
J0850-0318 & 59883 & 17 & 166$\pm$12 & 173$\pm$21 & 25 & $<$150 & {} & 35 & $<$210 & {} \\
J0850-0318 & 59904 & 15 & 168$\pm$14 & 264$\pm$31 & 20 & $<$120 & {} & 30 & $<$180 & {} \\
J1032-2707 & 59862 & 22 & 453$\pm$35 & 1047$\pm$81 & 35 & 243$\pm$28 & 473$\pm$65 & 40 & $<$240 & {} \\
J1032-2707 & 59903 & 15 & 585$\pm$33 & 1031$\pm$70 & 22 & 340$\pm$23 & 663$\pm$55 & 30 & 243$\pm$20 & 249$\pm$37\\
J1044-1826 & 59862 & 20 & 704$\pm$40 & 680$\pm$49 & 33 & 461$\pm$30 & 493$\pm$43 & 37 & 306$\pm$36 & 406$\pm$79\\
J1044-1826 & 59903 & 15 & 638$\pm$33 & 636$\pm$36 & 22 & 407$\pm$26 & 387$\pm$36 & 30 & 280$\pm$24 & 205$\pm$32\\
J1147-2145 & 59903 & 15 & 703$\pm$36 & 712$\pm$41 & 22 & 440$\pm$29 & 445$\pm$43 & 25 & 269$\pm$27 & 323$\pm$52\\
J2021-2235 & 59900 & 15 & 2653$\pm$133 & 2818$\pm$145 & 35 & 1543$\pm$91 & 1790$\pm$134 & 32 & 961$\pm$65 & 1160$\pm$109\\
J2244-1822 & 59884 & 15 & 122$\pm$16 & 158$\pm$33 & 25 & $<$150 & {} & 30 & $<$180 & {} \\
J2358-1028 & 59884 & 15 & 97$\pm$11 & 83$\pm$19 & 25 & $<$150 & {} & 30 & $<$180 & {} \\
\hline
    \end{tabular}
    \tablefoot{Columns: (1) Source name; (2) modified Julian date (MJD) of the observation; (3) rms of the map {at 15 GHz}; (4) peak flux density {at 15 GHz} ($\mu$Jy beam$^{-1}$); (5) integrated flux density {at 15 GHz} ($\mu$Jy); (6) rms of the map {at 22 GHz}; (7) peak flux density {at 22 GHz} ($\mu$Jy beam$^{-1}$); (8) integrated flux density {at 22 GHz} ($\mu$Jy); (9) rms of the map {at 33 GHz}; (10) peak flux density {at 33 GHz} ($\mu$Jy beam$^{-1}$); (11) integrated flux density {at 33 GHz} ($\mu$Jy). The * sign indicates a technical problem with the {22 GHz} observations of J0413-0050, where no data were taken.}
    \label{tab:data}
\end{table*}
\begin{table*}[!t]
    \centering
   \footnotesize
   \caption{Multi-epoch measurements of sources with extended morphology.}
    \begin{tabular}{l c c c c c c c c c}
    \hline
       Source & MJD & F$^c_{Ku}$ & F$^e_{Ku}$ & F$^c_{K}$ & F$^e_{K}$ & F$^c_{Ka}$ & F$^e_{Ka}$ & $\alpha^c_\nu$ & $\alpha^e_\nu$ \\
       \hline\hline
    J0452-2953 & 59872 & 1193$\pm$53 & 223$\pm$32 & 692$\pm$118 & 164$\pm$42 & 321$\pm$91 & 345$\pm$115 & -1.54$\pm$0.33 & 0.26$\pm$0.43 \\
    J0452-2953 & 59882 & 1194$\pm$54 & 95$\pm$29 & 511$\pm$60 & 362$\pm$63 & 376$\pm$83 & 289$\pm$100 & -1.78$\pm$0.22 & 1.54$\pm$0.58 \\
    J1032-2707 & 59862 & 423$\pm$45 & 645$\pm$60 & 306$\pm$91 & 403$\pm$114 & $<240$ & $<240$ & -0.85$\pm$0.82 & -1.23$\pm$0.78 \\
    \hline
    \end{tabular}
     \tablefoot{Columns: (1) Source name; (2) modified Julian date (MJD) of the observation; (3) integrated flux of the core component {at 15 GHz} ($\mu$Jy); (4) integrated flux of the extended emission {at 15 GHz} ($\mu$Jy); (5)  integrated flux of the core component {at 22 GHz} ($\mu$Jy); (6) integrated flux of the extended emission {at 22 GHz} ($\mu$Jy); (7) integrated flux of the core component {at 33 GHz} ($\mu$Jy); (8) integrated flux of the extended emission {at 33 GHz} ($\mu$Jy); (9) spectral index of the core component; (10) spectral index of the extended emission.}
    \label{tab:2dgauss}
\end{table*}

\section{Results}
\subsection{Radio morphology}
\label{sec:radio_morph}

At the frequencies of our observations, all of the sources but three appear point-like, with no signs of extended emission. The exceptions are J0452-2953, J0622-2317, and J1032-2707, which show signs of extended morphology, shown in Fig.~\ref{fig:J0452_map}, ~\ref{fig:J0622_map}, ~\ref{fig:J1032_map}. 

In J0452-2953, we can see elongated emission $\sim$23$^\circ$ in the south-east direction, which extends for $\sim$3", that is $\sim${13}~kpc {in projected size}. The same feature is visible {at 22 and 33 GHz}. For this reason, we decided to fit the source with two Gaussians, the first one representing the core, and the second the one-sided extended emission. We did the same for all the frequencies, and for both observing epochs (MJD 59872 and 59882). The results are reported in Table~\ref{tab:2dgauss}. The observations were taken ten days apart, and the integrated flux of the core and its spectral index remain constant over this interval. The steepness of the emission in the core may indicate that it originates from star formation in the nucleus, although it is possible that nuclear outflows are partially responsible for this emission. It is worth noting that the optical image of the galaxy, shown in Fig.~\ref{fig:J0452_host}, appears to be significantly elongated. Indeed, at a closer inspection, the galaxy shows two emission peaks 1.95" apart in the Pan-STARRS image ({8.4} kpc assuming that they are at the same distance), likely indicating that it is an interacting system. The radio emission seems to be associated with the south-east component. 

The extended flux instead seemingly shows some fast variability, especially {at 15 and 22 GHz}, and an inverted spectrum. Given the faintness of the emission, it is possible that the variability is only due to a non-perfect 2D fitting of the component, as it is highly unlikely to observe fast variability on a kpc-scale emission. Regarding the inverted spectrum, it may suggest the extended emission could originate in relativistic jets interacting with the interstellar medium (ISM), where the electrons are re-accelerated by this interaction. A similar behavior has already been observed in a couple more jetted NLS1s \citep{Jarvela22, Vietri22}, and can in principle affect also the optical emission lines \citep{Hon23, Dallabarba25}. It is worth noting that the axis of the radio emission and that of the two nuclei are closely aligned, suggesting a possible connection between the two phenomena, {although we cannot rule out that this is merely a projection effect}. Deeper observations are needed to reach more precise conclusions on the nature of this object.  

\begin{figure}
    \centering
    \includegraphics[width=\columnwidth]{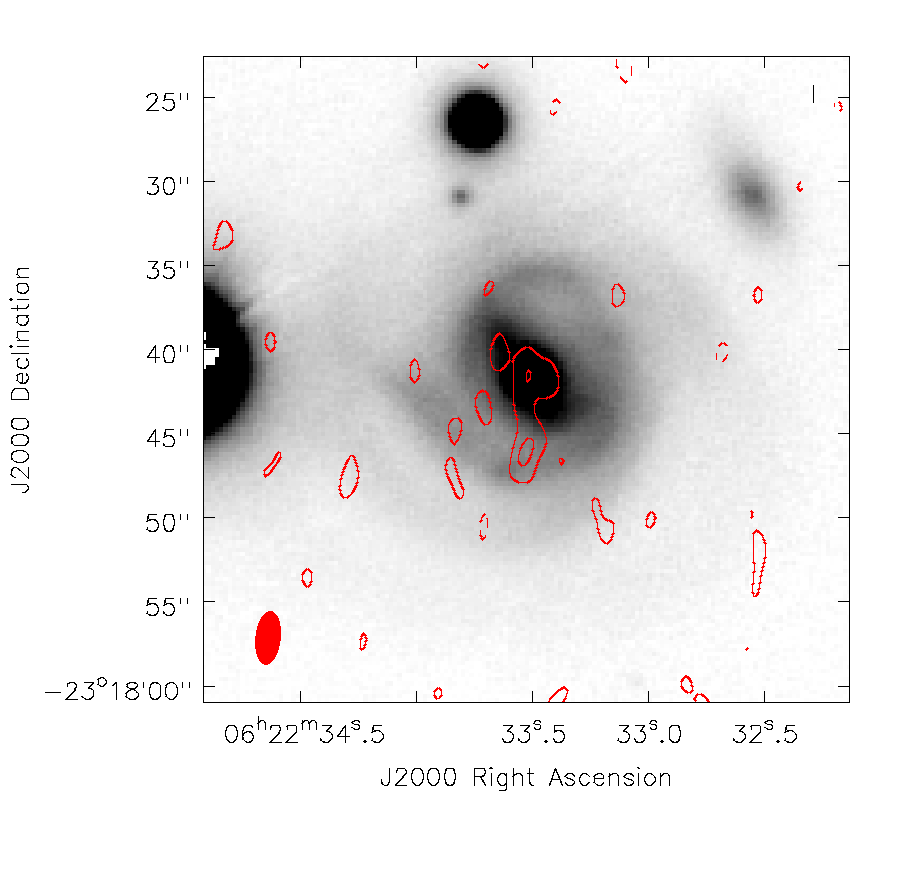}
    \caption{In red the contours of J0622-2317 as in Fig.~\ref{fig:J0622_map}, overlapped with the i-band image of its host galaxy extracted from Pan-STARRS.}
    \label{fig:J0622_host}
\end{figure}

The source J0622-2317 was observed on MJD 59898 and 59899, but it was detected only {at 15 GHz} on the first date. The morphology is extended in the South direction for $\sim$6.5" ($\sim${4.8} kpc), and it also shows significant diffuse emission which cannot be fitted with two Gaussian components. The source was already observed by \citet{Chen20}, who only found a point-like compact morphology at 5 GHz with the JVLA. Such a spread-out morphology may originate from star formation. The spiral morphology of the host, shown in Fig.~\ref{fig:J0622_host}, is also supporting this hypothesis, as star formation activity could be ongoing both in the nucleus and in the spiral arms. Its integrated luminosity, calculated on the 3$\sigma$ contour of the map, is rather low at $\log L =$ 37.95$\pm$0.06 \ergs, and it also supports the star formation origin of the radio. 

Finally, we analyzed J1032-2707, whose {15-GHz} map is shown in Fig.~\ref{fig:J1032_map}. The source was observed twice, on MJD 59862 and MJD 59903. In the first observation, the object clearly shows a double component {at 15 GHz}, as it does {at 22 GHz}, while {at 33 GHz} it was not detected. The emission {at 15 GHz} is extended for $\sim$5.2" ($\sim${7.1} kpc {of projected size}) approximately 33$^\circ$ North-East of the core. The two-point spectral index for the component corresponding to the optical nucleus is -0.85$\pm$0.82, while the other component has a spectral index of -1.23$\pm$0.78. Both were calculated using the integrated flux of the two Gaussian components. On the second observation, unfortunately, the beam is aligned almost exactly with the extended emission, and it is therefore preventing us from separating the two components and calculating two separate spectral indexes. It is worth noting that the extended emission is, at both frequencies, brighter than the core, with integrated luminosity {at 15 GHz} of $\log L = 39.08\pm0.05$ \ergs, compared to 38.89$\pm$0.05 \ergs\ for the core. Its spectral index is rather steep, and it is not clear if this emission corresponds to a relativistic jet, or if it is due to intense star formation, although the latter seems exceedingly bright at these frequencies.

\subsection{Spectral shape}
\label{sec:spectra}
Out of 26 objects, three do not have enough data to properly fit their spectrum. J0413-0050 was detected only at two frequencies. We calculate that its two-point spectral index of the peak flux density of this source is -0.42$\pm$0.13, that is between a flat- and a steep-spectrum source (threshold -0.5, see \citealp{Foschini15}). J2244-1822 and J2358-1029 have only been detected {at 15 GHz} by the JVLA, suggesting that their spectral index is negative. 

In four cases, the spectrum can be better reproduced with a log parabola. As previously mentioned, this could potentially suggest the presence of a curvature, which in turn could indicate some form of absorption, possibly SSA, at low frequencies. This feature could indicate the presence of a young relativistic jet, whose spectra indeed are peaked \citep{Fanti95}. We tried to extrapolate the position of the peak in the spectrum. Two of them peak in the MHz range, while the other two peak in the kHz range. However, it is worth noting that only in one source, J0447-0508, the peak position is relatively well constrained, while for the others, the uncertainty is large. Let us assume that the spectral peak is related to the jet size, as described in Eq. 4 of \citet{Odea98}. If this is the case, the sources peaking in the kHz range would have a jet of Mpc scale, which is not seen at any frequency or scale. Only in the case of J0447-0508 the peak frequency would correspond to a jet possibly confined within the host galaxy, $\sim$2.97 kpc. However, the scale for this source is 0.846 kpc/". The corresponding linear size {for a 90$^\circ$ inclination would be} 3.5" on the sky{, but it could be smaller if the jet had a lower inclination}. No sign of extended emission is visible in our maps. {Therefore, either there is a jet with a rather small inclination, but we would then start seeing effects from relativistic beaming such as significant variability. Otherwise, the spectral} curvature, if at all there, does not seem to imply that a jet is present. 

One source is particularly interesting, J0239-1118. The source was detected by \citet{Chen20} at 5 GHz, showing no particular features, but the high frequency observations revealed a strongly inverted spectrum, with $\alpha_\nu = 0.28\pm0.04$. The peak of the emission is therefore at higher frequencies, and due to the lack of data, we could not constrain its position. In principle, an inverted spectrum at high frequency can originate from free-free absorption (FFA, \citealp{Condon92}). However, its spectral index in this case is expected to be much more inverted than what we observe, and on average, more inverted than SSA. Therefore, we suggest that this source is a high-frequency peaker (HFP, \citealp{Orienti09}), that is, a newborn relativistic jet growing quickly.  

Another object in the sample, J0203-1247, has a spectral index of $-0.35\pm0.05$, consistent with a classification as a flat-spectrum object. Its luminosity at 5 GHz was already calculated by \citet{Chen20}, who reported a value of 1.3$\times10^{38}$ \ergs. This is much lower compared to the typical value of jetted NLS1s \citep{Berton18a}, therefore, the origin of this emission in this case may reside in a non-relativistic outflow produced by the nucleus, and not a relativistic jet. 

Finally, the remaining 18 objects all show a negative slope, with values ranging between -0.5 and -0.94. These numbers, along with their point-like morphology in all of our observations, are all consistent with optically thin synchrotron emission. Its origin could be a jet, as is likely in J0452-2953, or from outflows, a jet base, or star formation.





\subsection{Radio luminosity}
We calculated the radio luminosity from the flux densities reported in Table~\ref{tab:data}, following
\begin{equation}
L_{\nu} = 4 \pi D_L^2\, S_{\nu_\mathrm{obs}}\, (1+z)^{-(\alpha + 1)}\; , 
\end{equation}
where $D_L$ is the luminosity distance, $z$ the source's redshift, and $\alpha$ is the spectral index, defined as $S_\nu \propto \nu^\alpha$. The spectral index was calculated directly on the data using the nearest frequency point. When this was not possible, for example, for upper limits, we used $\alpha = -1$. We report all the luminosities in Table~\ref{tab:luminosity}. In \citet{Berton18a}, all the sources which most likely harbor a relativistic jet had a luminosity larger than $10^{40}$ \ergs\ at 5 GHz. Assuming the typical spectral index of synchrotron radiation, i.e. -0.7, this corresponds to a logarithmic luminosity of 39.67, 39.56, and 39.44 {at 15, 22, and 33 GHz}, respectively. Six of our sources exceed these thresholds in at least one frequency. This could suggest the presence of jets in these sources, but it is not so straightforward since strong star formation can also reach similar luminosities \citep{Caccianiga15}. Finally, we checked for significant variability among sources with multiple observations. However, we did not find anything above 3$\sigma$, only some marginal differences at most within 2$\sigma$. 

\section{Discussion}
\subsection{The high frequency spectrum of NLS1s}
In most sources, the main contribution to the radio emission probably comes from the AGN itself, that is, the synchrotron emission produced by non-relativistic outflows, or perhaps by the intense circumnuclear star formation that is often seen in NLS1s \citep{Sani10, Winkel22, Winkel23}. In principle, radio emission can also be produced by the accretion disk corona, but that typically produces a flat spectrum that we do not observe here \citep{Raginski16}.  {The accretion disk itself could be contributing to the radio spectrum by means of synchrotron or free-free emission \citep{Abramowicz13}. Free-free emission, in particular, could be detectable at high radio frequency, with a characteristic slope of $\alpha_\nu$ = 0.1 \citep{Condon92}. However, our spectral indexes are typically steeper than that, and are more indicative of synchrotron emission. Furthermore, the peak of the accretion disk usually falls in the ultraviolet band \citep[e.g.,][]{Czerny87, Panda18a, Ferland20, Garnica25}, so if the disk contribution is present in our observations it is probably rather small.} Other potential sources of steep-spectrum radio emission are a jet base \citep[e.g.,][]{Giroletti09}, failed relativistic jets \citep{Ghisellini04}, or fully developed small-scale relativistic jets seen at large angles. However, when we compare their luminosity to the sources studied by \citet{Berton18a}, our objects are within the typical range of nonjetted sources, although we cannot completely rule out the presence of relativistic jets in them. For example, in J1032-2707 the luminosity is rather low, around $10^{39}$ \ergs\, but its morphology shown in Fig.~\ref{fig:J1032_map} is reminiscent of a core-jet system. It is possible that this extended emission is a non-relativistic outflow, but only additional observations could confirm this scenario.

It is also worth noting that in our sources we do not see the so-called high-frequency excess (HFE) observed by \citet{Antonucci88} and \citet{Barvainis96}. With early VLA observations, they showed that several radio-quiet quasars or luminous Seyfert galaxies showed a steep spectrum at low frequencies (below 10 GHz), that turned flat or inverted above this frequency. Some objects showed an excess at even higher frequency (95 GHz), suggesting that the corona is responsible for this emission \citep{Behar15}, although this point is still debated \citep{Doi16a}. A similar behavior was also observed in low-luminosity AGN, and in that case, the nature of the inverted spectrum was attributed to an advection-dominated accretion flow \citep{Doi05}. Interestingly, some of the sources studied by \citet{Behar15} are well-known NLS1s (Ark 564, Mrk 766, and also NGC 5506, the first obscured NLS1 ever discovered \citealp{Nagar02}). This suggests the presence of diverse properties within the NLS1 class, which remain to be understood. 

Finally, these radio spectra do not seem to be strongly variable, which is also consistent with the outflow scenario. In particular, we do not observe any sign of the extreme variability detected by MRO at 37 GHz and by OVRO at 15 GHz \citep{Lahteenmaki18, Jarvela24}. However, it is worth noting that the timescale for this new type of variability could be shorter than the cadence between our observations. Already in \citet{Jarvela24} they found that the $e$-folding time of these flares can be as fast as one day, but more recent observations indicate that it could be even shorter (Crepaldi et al. in prep., J\"arvel\"a et al. in prep.). Therefore, it is not very surprising that the JVLA did not catch any of these flaring events. With our current knowledge, only dedicated monitoring at high frequency could find new examples of this intriguing phenomenon. 

\subsection{New jetted NLS1s?}
\begin{figure}
    \centering
    \includegraphics[width=\columnwidth]{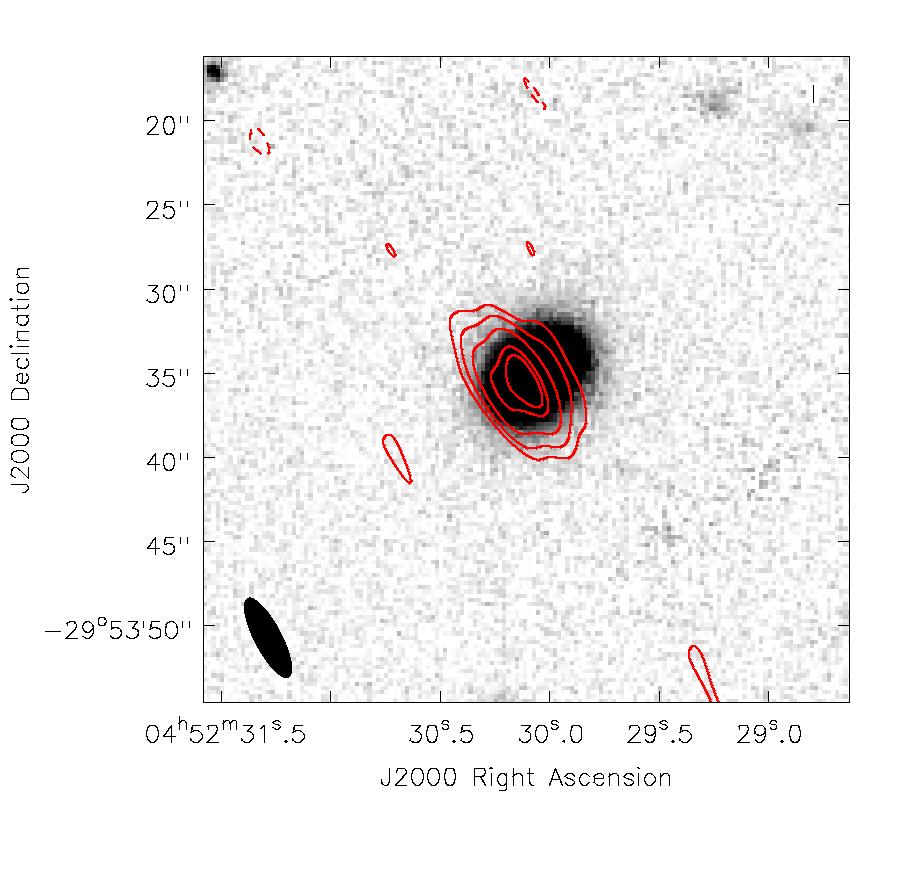}
    \caption{In red the contours of J0452-2953 as in Fig.~\ref{fig:J0452_map}, overlapped with the i-band image of its host galaxy extracted from Pan-STARRS.}
    \label{fig:J0452_host}
\end{figure}

It is evident from our work and that of many others \citep{Berton18a, Chen20, Chen22, Chen23, Jarvela22} that identifying the nature of the radio emission in NLS1s is extremely challenging. Spectral indexes and morphology provide some indication, but do not paint the whole picture. Multiwavelength and multiscale observations are necessary to give more precise indications on the nature of these sources. It is also worth noting that our data, taken {at 15, 22, and 33 GHz} with the JVLA C configuration, can only study relatively small spatial scales, i.e. a few kpc. Some recent studies showed that some relativistic jets in NLS1 can have a very large size \citep{Rakshit18, Vietri22, Chen24}, and can therefore be resolved out at the spatial resolution we are currently observing \citep{Umayal25}. This, however, should not be the case for our objects, since the low-frequency points do not show any flux excess suggestive of extended emission.  

However, with this work, we likely identified two new jetted NLS1s, J0239-1118, and J0452-2953. The latter is the most luminous source in our sample, with an integrated luminosity {at 33 GHz} of $\sim5\times10^{40}$ \ergs, comparable to that of other known jetted NLS1s \citep{Berton18a}. It also shows an elongated emission that extends for {13} kpc {in projected size} toward south-east with an inverted spectrum. We attribute these properties to a possible relativistic jet interacting with the ISM. As shown in Fig.~\ref{fig:J0452_host}, the radio emission {seems} almost entirely confined within the host galaxy, so jet/ISM interaction is definitely possible{, although this alignment may be coincidental and due just to projection effects}. Other possible explanations for the inverted spectrum are forms of absorption, FFA or SSA, but they appear less likely at such a distance from the nuclear region. If the emission is due to a relativistic jet{, considering its projected size as a lower limit and} assuming a propagation velocity of $\sim$0.5$c$ \citep{Giroletti09a}, its age should be $>$7.8$\times10^4$ years, within the typical range for NLS1s \citep{Czerny09}. 

As for J0239-1118, as we already mentioned, it could be an HFP. These objects belong to the larger class of peaked sources (PS, \citealp{Odea21}), which is a mixed bag of different objects sharing one common observational property, that is, a peaked radio spectrum. The peak is due to significant SSA toward lower frequencies, which indicates that PS are typically small-scale relativistic jets still confined within their host galaxy. The PS class includes a number of frustrated jets that are not powerful enough to escape their host. At the same time, it also includes genuinely young sources such as compact steep-spectrum sources (CSS), gigahertz-peaked sources (GPS), and HFPs, in which the small size of the jet corresponds to young age. Since there is an anticorrelation between the jet age and the peak frequency \citep{Odea97}, HFPs are the youngest among all jets, so young that their spectra evolve quickly on an observable timescale. We therefore expect to observe significant variability in J0239-1118 within the next few years, which will allow us to constrain the peak position and confirm its nature as a young relativistic jet. 

The presence of such a young jet in an NLS1 should not come as a surprise. The first suggestion of a connection between NLS1s and PS came already in the early 2000s, when some authors noted that a jetted NLS1 presented a peaked spectrum \citep{Oshlack01, Schulz15}. In the following years, it was suggested that some PSs, in particular the class of low-luminosity compact sources \citep{Kunert10a} may be part of the parent population of gamma-ray NLS1s, since their radio luminosity functions \citep{Berton16c} and host galaxy properties \citep{Vietri24} match each other. Finding an HFP in an NLS1 strengthens this view, proving once again that part of the parent population of gamma-ray emitting NLS1s can indeed be found among PS \citep{Berton17, Berton21c}. 

Sources like this may be relatively common at high redshift. Indeed, to form the massive seeds that we observe in the early Universe \citep{Inayoshi20}, relativistic jets can be instrumental in draining angular momentum and allowing for super-Eddington accretion \citep[e.g.,][]{Volonteri21}. Recently, one blazar was identified at the epoch of reionization \citep{Banados25}, implying that a number of parent population sources should exist. Assuming that these objects look like J0239-1118, we can extrapolate the expected flux density at 3.0 GHz, where, for example, VLASS is currently observing. Assuming a spectral index of $\alpha_\nu$ = 0.28, and extrapolating at $z$ = 6, the flux density would be 1.96 $\mu$Jy, which is well below the detection limit of any present-day survey. This estimate shows why only a handful of misaligned jets have been identified at the epoch of reionization \citep{Banados21, Endsley23}. However, these sources could become detectable with future instrumentation such as the Square Kilometre Array (SKA). 

Last, it is worth mentioning the source J2021-2235, a.k.a. IRAS 20181-2244. This NLS1 is hosted by a late-type galaxy interacting with a satellite \citep{Berton19a}, and it was identified as a jetted object by \citet{Komossa06}. Based on its infrared emission, it seems to have a remarkably high star formation of 300 M$_\odot yr^{-1}$, which is responsible for a very significant fraction of its radio emission \citep{Caccianiga15}. Its relativistic jet appears unresolved in our data as it was at 5 GHz \citep{Chen20}, implying that it is either resolved out, {genuinely} very small, {or it has a small projected size due to a low inclination}. It is worth noting that J2021-2235 is the second most luminous source in our sample ($\sim4\times10^{40}$ \ergs\ integrated {at 15 GHz}), and this suggests that the jet contribution plays a significant role also at these frequencies{, possibly due to some limited contribution of relativistic beaming if the viewing angle is small.}

\section{Conclusions}
In this paper, we report on our new JVLA observations of a sample of 50 NLS1s at 15, 22, and 33~GHz. Approximately half of the sources, 26, were detected at least at one frequency. We studied their radio properties by adding available data from the literature. Their spectrum is almost always steep and can be reproduced by a power law. The emission possibly originates in radiation pressure-driven non-relativistic outflows or, less likely, in circumnuclear star formation. The origin of these seemingly common outflows may lie in the strong radiation pressure produced by the high accretion typical of NLS1s. A couple of sources may harbor small-scale relativistic jets, and one in particular may be a high-frequency peaker. It is possible that more relativistic jets are present, but we cannot confirm their presence with the available data. These results confirm the complexity and diverse nature of the NLS1 class. Additional multiwavelength studies are necessary to fully characterize these sources, and understand their role in the AGN life cycle. 



\begin{table*}[]
    \centering
    \footnotesize
    \caption{Parameters of linear and parabolic best fit for sources.}
    \begin{tabular}{l c c c c c c }
    Short name & Slope $\alpha_\nu$ & Parabola peak & $\chi_l$ & $\chi_p$ & F p-value & Preferred model \\
    \hline\hline
    J0000-0541 & -0.94$\pm$0.04 & (681$\pm$533) kHz & 169.09 & 38.70 & 1.27$\times10^{-2}$ & Log parabola \\
    J0022-1039 & -0.88$\pm$0.07 & * & 2163.65 & - & - & Power law \\
    J0203-1247 & -0.35$\pm$0.05 & * & 42.93 & - & - & Power law \\
    J0212-0201 & -0.68$\pm$0.06 & - & 0.81 & - & - & Power law \\
    J0213-0551 & -0.47$\pm$0.04 & - & 0.41 & - & - & Power law \\
    J0230-0859 & -0.75$\pm$0.03 & * & 100.07 & - & - & Power law \\
    J0239-1118 & +0.28$\pm$0.04 & - & 45.03 & - & - & Power law \\
    J0400-2500 & -0.91$\pm$0.10 & (250$\pm$183) MHz & 1003.23 & 528.83 & 1.27$\times10^{-1}$ & Power law \\
    J0413-0050 & & & & & & Not enough points \\
    J0422-1854 & -0.87$\pm$0.05 & (18$\pm$17) MHz & 472.39 & 173.13 & 2.10$\times10^{-2}$ & Log parabola \\
    J0436-1022 & -0.77$\pm$0.04 & * & 15303.97 & - & - & Power law \\
    J0447-0508 & -0.75$\pm$0.06 & (304$\pm$105) MHz & 1496.13 & 161.35 & 5.19$\times10^{-4}$ & Log parabola \\
    J0452-2953 & -1.05$\pm$0.10 & * & 1657.67 & - & - & Power law \\
    J0549-2425 & -0.80$\pm$0.03 & (707$\pm$576) KHz & 235.97 & 25.21 & 6.27$\times10^{-4}$ & Log parabola \\
    J0622-2317 & -1.29$\pm$0.13 & - & 55.83 & - & - & Power law \\
    J0820-1741 & -0.84$\pm$0.13 & - & 16.64 & - & - & Power law \\
    J0842-0349 & -0.79$\pm$0.04 & - & 0.36 & - & - & Power law \\
    J0846-1214 & -0.90$\pm$0.02 & (3.16$\pm$2.24) Hz & 18362.57 & 7549.26 & 7.21$\times10^{-2}$ & Power law \\
    J0849-2351 & -0.81$\pm$0.05 & * & 56.06 & - & - & Power law \\
    J0850-0318 & -0.74$\pm$0.09 & * & 495.52 & - & - & Power law \\
    J1032-2707 & -0.94$\pm$0.07 & * & 623.56 & - & - & Power law \\
    J1044-1826 & -0.82$\pm$0.02 & * & 56.83 & - & - & Power law \\
    J1147-2145 & -0.87$\pm$0.05 & (6.97$\pm$2.53) MHz & 185.45 & 58.96 & 7.71$\times10^{-2}$ & Power law \\
    J2021-2235 & -0.88$\pm$0.07 & * & 2163.65 & - & - & Power law \\
    J2244-1822 & & & & & & Not enough points \\
    J2358-1029 & & & & & & Not enough points \\
    \hline
    \end{tabular}
    \tablefoot{Columns: (1) Source name; (2) slope of the linear fit, that is spectral index $\alpha_\nu$; (3) peak of the parabolic fit. The ``*" symbol indicates that the log parabola coefficients are consistent with a power law, the ``-" indicates that the fit did not converge due to too few points; (4) chi-squared of the linear fit; (5) chi-squared of the parabolic fit; (6) p-value of the F test, (7) preferred model based on F test. We fix the threshold at p-value$<$5$\times10^{-2}$.}
    \label{tab:spind}
\end{table*}

\begin{acknowledgements}
We thank the anonymous referee for constructive comments that helped improving the paper. The authors are grateful to Dr. Elisa Garro for useful discussion on the nature of young relativistic jets. M.B. and E.J. acknowledge the support of the ESO Chile Visitor Program. A.J.G. and C.P. acknowledge the ESO Science Support Discretionary Fund 2025. I.V. acknowledges the support of the Swedish Cultural Foundation in Finland and the financial support from the Visitor and Mobility program of the Finnish Centre for Astronomy with ESO (FINCA). S.P. is supported by the international Gemini Observatory, a program of NSF NOIRLab, which is managed by the Association of Universities for Research in Astronomy (AURA) under a cooperative agreement with the U.S. National Science Foundation, on behalf of the Gemini partnership of Argentina, Brazil, Canada, Chile, the Republic of Korea, and the United States of America. The National Radio Astronomy Observatory and Green Bank Observatory are facilities of the U.S. National Science Foundation operated under cooperative agreement by Associated Universities, Inc. This research has made use of the NASA/IPAC Extragalactic Database (NED), which is operated by the Jet Propulsion Laboratory, California Institute of Technology, under contract with the National Aeronautics and Space Administration. This research has made use of the SIMBAD database, operated at CDS, Strasbourg, France. This research has made use of the VizieR catalog access tool, CDS, Strasbourg, France. The National Radio Astronomy Observatory is a facility of the National Science Foundation operated under cooperative agreement by Associated Universities, Inc. This research has made use of the CIRADA cutout service at URL cutouts.cirada.ca, operated by the Canadian Initiative for Radio Astronomy Data Analysis (CIRADA). CIRADA is funded by a grant from the Canada Foundation for Innovation 2017 Innovation Fund (Project 35999), as well as by the Provinces of Ontario, British Columbia, Alberta, Manitoba and Quebec, in collaboration with the National Research Council of Canada, the US National Radio Astronomy Observatory and Australia’s Commonwealth Scientific and Industrial Research Organisation. The National Radio Astronomy Observatory is a facility of the National Science Foundation operated under cooperative agreement by Associated Universities, Inc. LOFAR data products were provided by the LOFAR Surveys Key Science project (LSKSP; https://lofar-surveys.org/) and were derived from observations with the International LOFAR Telescope (ILT). LOFAR \citep{Vanhaarlem2013} is the Low Frequency Array designed and constructed by ASTRON. It has observing, data processing, and data storage facilities in several countries, which are owned by various parties (each with their own funding sources), and which are collectively operated by the ILT foundation under a joint scientific policy. The efforts of the LSKSP have benefited from funding from the European Research Council, NOVA, NWO, CNRS-INSU, the SURF Co-operative, the UK Science and Technology Funding Council and the Jülich Supercomputing Centre. The Pan-STARRS1 Surveys (PS1) and the PS1 public science archive have been made possible through contributions by the Institute for Astronomy, the University of Hawaii, the Pan-STARRS Project Office, the Max-Planck Society and its participating institutes, the Max Planck Institute for Astronomy, Heidelberg and the Max Planck Institute for Extraterrestrial Physics, Garching, The Johns Hopkins University, Durham University, the University of Edinburgh, the Queen’s University Belfast, the Harvard-Smithsonian Center for Astrophysics, the Las Cumbres Observatory Global Telescope Network Incorporated, the National
Central University of Taiwan, the Space Telescope Science Institute, the National Aeronautics and Space Administration under Grant No. NNX08AR22G issued through the Planetary Science Division of the NASA Science Mission Directorate, the National Science Foundation Grant No. AST-1238877, the University of Maryland, Eotvos Lorand University (ELTE), the Los Alamos National Laboratory, and the Gordon and Betty Moore Foundation. 

\end{acknowledgements}

%
\bibliographystyle{aa.bst} 
\bibliography{biblio} 
\newpage
\begin{appendix}
\section{Flux measurements}
\begin{table*}[!h]
    \centering
    \footnotesize
    \caption{Upper limits for sources with no detection.}
    \begin{tabular}{l c c c c c c c}
Name & MJD & rms (Ku) & S$_p$ (Ku) & rms (K) & S$_p$ (K) & rms (Ka) & S$_p$ (Ka) \\
\hline\hline
J0015-1509 & 59904 & 15 & $<$90 & 18 & $<$108 & 25 & $<$150 \\
J0021-2050 & 59904 & 15 & $<$90 & 18 & $<$108 & 25 & $<$150 \\
J0030-2028 & 59904 & 15 & $<$90 & 18 & $<$108 & 25 & $<$150 \\
J0043-1655 & 59904 & 15 & $<$90 & 18 & $<$108 & 25 & $<$150 \\
J0200-0845 & 59863 & 20 & $<$120 & 30 & $<$180 & 40 & $<$240 \\
J0200-0845 & 59902 & 13 & $<$78 & 19 & $<$114 & 25 & $<$150 \\
J0212-0737 & 59863 & 19 & $<$114 & 30 & $<$180 & 41 & $<$246 \\
J0212-0737 & 59902 & 13 & $<$78 & 25 & $<$150 & 25 & $<$150 \\
J0420-0530 & 59872 & 20 & $<$120 & 20 & $<$120 & 30 & $<$180 \\
J0420-0530 & 59882 & 20 & $<$120 & 30 & $<$180 & 30 & $<$180 \\
J0435-1643 & 59872 & 13 & $<$78 & 22 & $<$132 & 33 & $<$198 \\
J0435-1643 & 59882 & 15 & $<$90 & 25 & $<$150 & 35 & $<$210 \\
J0447-0403 & 59872 & 14 & $<$84 & 20 & $<$120 & 30 & $<$180 \\
J0447-0403 & 59882 & 14 & $<$84 & 25 & $<$150 & 30 & $<$180 \\
J0455-1456 & 59872 & 12 & $<$72 & 21 & $<$126 & 31 & $<$186 \\
J0455-1456 & 59882 & 15 & $<$90 & 25 & $<$150 & 36 & $<$216 \\
J0845-0732 & 59883 & 18 & $<$108 & 27 & $<$162 & 35 & $<$210 \\
J0845-0732 & 59904 & 15 & $<$90 & 20 & $<$120 & 30 & $<$180 \\
J1014-0418 & 59862 & 20 & $<$120 & 35 & $<$210 & 40 & $<$240 \\
J1014-0418 & 59903 & 15 & $<$90 & 22 & $<$132 & 30 & $<$180 \\
J1015-1652 & 59862 & 24 & $<$144 & 29 & $<$174 & 42 & $<$252 \\
J1015-1652 & 59903 & 15 & $<$90 & 22 & $<$132 & 30 & $<$180 \\
J1032-1609 & 59862 & 20 & $<$120 & 30 & $<$180 & 40 & $<$240 \\
J1032-1609 & 59903 & 15 & $<$90 & 22 & $<$132 & 30 & $<$180 \\
J1057-0805 & 59862 & 25 & $<$150 & 33 & $<$198 & 40 & $<$240 \\
J1057-0805 & 59903 & 25 & $<$150 & 22 & $<$132 & 30 & $<$180 \\
J2115-1417 & 59900 & 900 & $<$5400 & 150 & $<$900 & 33 & $<$198 \\
J2136-0116 & 59900 & 12 & $<$72 & 17 & $<$102 & 25 & $<$150 \\
J2137-1112 & 59900 & 13 & $<$78 & 17 & $<$102 & 25 & $<$150 \\
J2143-2958 & 59900 & 13 & $<$78 & 19 & $<$114 & 25 & $<$150 \\
J2155-1210 & 59900 & 13 & $<$78 & 19 & $<$114 & 25 & $<$150 \\
J2207-2824 & 59884 & 17 & $<$102 & 30 & $<$180 & 35 & $<$210 \\
J2229-1401 & 59884 & 15 & $<$90 & 25 & $<$150 & 32 & $<$192 \\
J2250-1152 & 59884 & 15 & $<$90 & 23 & $<$138 & 30 & $<$180 \\
J2311-2022 & 59884 & 15 & $<$90 & 25 & $<$150 & 35 & $<$210 \\
\hline
    \end{tabular}
    \tablefoot{Columns: (1) Source name; (2) modified Julian date (MJD) of the observation; (3) rms of the map {at 15 GHz}; (4) upper limit of the peak flux density {at 15 GHz} ($\mu$Jy beam$^{-1}$); (5) rms of the map {at 22 GHz}; (6) upper limit of the peak flux density {at 22 GHz} ($\mu$Jy beam$^{-1}$); (7) rms of the map {at 33 GHz}; (8) upper limit of the peak flux density {at 33 GHz} ($\mu$Jy beam$^{-1}$).}
    \label{tab:nodetection}
\end{table*}
\clearpage

\begin{sidewaystable*}[]
    \centering
    \footnotesize
    \caption{Archival radio data for the sources with at least one JVLA detection.}
    \scalebox{0.9}{
    \begin{tabular}{ccccccccccccccc}
        Source name & FIR$_{p}$ & FIR$_{i}$ & NVSS$_{i}$ & VL1$_{p}$ & VL1$_{i}$ & VL2$_{p}$ & VL2$_{i}$ & S5$_{i}$ & S5$_{p}$ & S9$_{p}$ & RACS$_{i}$ & RACS$_{p}$ & TGSS$_{i}$ & TGSS$_{p}$ \\ \hline \hline
        J0000-0541 & 2.98 & 2.49 & {} & 1.62 $\pm $0.12 & 2.07 $\pm $0.26 & 1.90 $\pm $0.15 & 2.36 $\pm $0.33 & 1164 $\pm $16 & 1058.5 $\pm $8.5 & {} & 15.25 $\pm $0.73 & 4.50 $\pm $0.29 & {} & {}   \\ 
        J0015-1509 & {} & {} & {} & {} & {} & {} & {} & {} & {} & {} & {} & {} & {} & {}   \\ 
        J0021-2050 & {} & {} & {} & {} & {} & {} & {} & {} & {} & {} & {} & {} & {} & {}   \\ 
        J0022-1039 & 1.66 & 1.40 & {} & {} & {} & {} & {} & 514 $\pm $27 & 425 $\pm $14 & {} & 2.78 $\pm $0.83 & 2.81 $\pm $0.48 & {} & {}   \\ 
        J0030-2028 & {} & {} & {} & {} & {} & {} & {} & {} & {} & {} & {} & {} & {} & {}   \\ 
        J0043-1655 & {} & {} & {} & {} & {} & {} & {} & {} & {} & {} & {} & {} & {} & {}   \\ 
        J0200-0845 & {} & {} & {} & {} & {} & {} & {} & {} & {} & {} & {} & {} & {} & {}   \\ 
        J0203-1247 & {} & {} & {} & {} & {} & {} & {} & 360 $\pm $10 & 314.9 $\pm $4.9 & {} & {} & {} & {} & {}   \\ 
        J0212-0201 & {} & {} & {} & {} & {} & {} & {} & 229.4 $\pm $8.8 & 223.3 $\pm $4.8 & {} & {} & {} & {} & {}   \\ 
        J0212-0737 & {} & {} & {} & {} & {} & {} & {} & {} & {} & {} & {} & {} & {} & {}   \\ 
        J0213-0551 & {} & {} & {} & {} & {} & {} & {} & 202 $\pm $9.6 & 166.5 $\pm $4.8 & {} & {} & {} & {} & {}   \\ 
        J0230-0859 & 1.32 & 1.17 & 2.40 $\pm $0.5 & 1.17 & {} & 1.45 $\pm $0.15 & 2.09 $\pm $0.33 & 1232 $\pm $25 & 1029 $\pm $13 & {} & 4.71 $\pm $0.62 & 4.27 $\pm $0.34 & {} & {}   \\ 
        J0239-1118 & {} & {} & {} & {} & {} & {} & {} & 327.7 $\pm $5.1 & 327.7 $\pm $5.1 & {} & {} & {} & {} & {}   \\ 
        J0400-2500 & {} & {} & 4.1 $\pm $0.6 & {} & {} & 0.92 $\pm $0.15 & 1.73 $\pm $0.41 & 1245 $\pm $14 & 1245 $\pm $14 & {} & 3.97 $\pm $0.43 & 4.03 $\pm $0.25 & {} & {}   \\ 
        J0413-0050 & {} & {} & {} & {} & {} & {} & {} & 161 $\pm $11 & 140.7 $\pm $5.9 & {} & {} & {} & {} & {}   \\ 
        J0420-0530 & {} & {} & {} & {} & {} & {} & {} & {} & {} & {} & {} & {} & {} & {}   \\ 
        J0422-1854 & {} & {} & 2.8 $\pm $0.5 & 1.11 $\pm $0.17 & 1.98 $\pm $0.45 & 1.43 $\pm $0.17 & 2.00 $\pm $0.38 & 1126 $\pm $21 & 1069 $\pm $11 & {} & 4.26 $\pm $0.66 & 3.57 $\pm $0.34 & {} & {}   \\ 
        J0435-1643 & {} & {} & {} & {} & {} & {} & {} & 141.1 $\pm $5.9 & 131.9 $\pm $2.8 & {} & {} & {} & {} & {}   \\ 
        J0436-1022 & {} & {} & 17 $\pm $0.7 & 5.47 $\pm $0.13 & 6.33 $\pm $0.26 & 5.51 $\pm $0.16 & 7.02 $\pm $0.34 & 4620 $\pm $220 & 4031 $\pm $96 & {} & 21.2 $\pm $2.2 & 18.85 $\pm $0.52 & 50.4 $\pm $7.4 & 29.9 $\pm $4.7   \\ 
        J0447-0403 & {} & {} & {} & {} & {} & {} & {} & 118 $\pm $12 & 42.3 $\pm $3.4 & {} & {} & {} & {} & {}   \\ 
        J0447-0508 & {} & {} & {} & 4.26 $\pm $0.13 & 4.94 $\pm $0.26 & 5.27 $\pm $0.14 & 5.77 $\pm $0.28 & 4040 $\pm $50 & 3600 $\pm $25 & {} & 89.0 $\pm $1.1 & 8.91 $\pm $0.35 & {} & {}   \\ 
        J0452-2953 & {} & {} & 9.5 $\pm $0.5 & 2.43 $\pm $0.18 & 4.85 $\pm $0.51 & 2.57 $\pm $0.13 & 6.14 $\pm $0.43 & 3373 $\pm $47 & 2564 $\pm $22 & {} & 19.1 $\pm $1.2 & 16.55 $\pm $0.62 & {} & {}   \\ 
        J0455-1456 & {} & {} & {} & {} & {} & {} & {} & 138.8 $\pm $5.2 & 136.1 $\pm $2.9 & {} & {} & {} & {} & {}   \\ 
        J0549-2425 & {} & {} & {} & {} & {} & {} & {} & 1560 $\pm $11 & 1560 $\pm $11 & {} & 5.09 $\pm $0.57 & 5.03 $\pm $0.33 & {} & {}   \\ 
        J0622-2317 & 4.3 & 0.5 & {} & {} & {} & {} & {} & 926 $\pm $59 & 281 $\pm $12 & {} & 5.9 $\pm $1.4 & 3.02 $\pm $0.49 & {} & {}   \\ 
        J0820-1741 & {} & {} & {} & {} & {} & {} & {} & {} & {} & {} & 1.73 $\pm $0.55 & 1.75 $\pm $0.32 & {} & {}   \\ 
        J0842-0349 & {} & {} & {} & {} & {} & {} & {} & 324 $\pm $18 & 269.4 $\pm $9.1 & {} & {} & {} & {} & {}   \\ 
        J0845-0732 & {} & {} & {} & {} & {} & {} & {} & 259 $\pm $26 & 127.7 $\pm $9 & {} & {} & {} & {} & {}   \\ 
        J0846-1214 & {} & {} & 15.3 $\pm $0.7 & 8.11 $\pm $0.17 & 8.81 $\pm $0.31 & 8.67 $\pm $0.13 & 9.52 $\pm $0.28 & 5310 $\pm $34 & 5237 $\pm $18 & {} & 30.8 $\pm $1.9 & 25.29 $\pm $0.49 & 86.2 $\pm $11.0 & 75.2 $\pm $8.6   \\ 
        J0849-2351 & {} & {} & {} & {} & {} & 1.19 $\pm $0.13 & 1.64 $\pm $0.29 & 921 $\pm $17 & 748.5 $\pm $7.4 & {} & 3.48 $\pm $0.53 & 3.26 $\pm $0.29 & {} & {}   \\ 
        J0850-0318 & {} & {} & {} & {} & {} & {} & {} & 426 $\pm $19 & 357.2 $\pm $9.7 & {} & 2.77 $\pm $0.81 & 2.10 $\pm $0.38 & {} & {}   \\ 
        J1014-0418 & {} & {} & {} & {} & {} & {} & {} & 380 $\pm $23 & 247.9 $\pm $9.8 & {} & 2.46 $\pm $0.61 & 2.12 $\pm $0.31 & {} & {}   \\ 
        J1015-1652 & {} & {} & {} & {} & {} & {} & {} & 73.3 $\pm $9.7 & 69.9 $\pm $4.9 & {} & {} & {} & {} & {}   \\ 
        J1032-1609 & {} & {} & {} & {} & {} & {} & {} & {} & {} & {} & {} & {} & {} & {}   \\ 
        J1032-2707 & {} & {} & 5.3 $\pm $0.6 & 1.10 $\pm $0.18 & 2.48 $\pm $0.57 & 1.35 & {} & 930 $\pm $60 & 930 $\pm $60 & 720 $\pm $50 & 7.73 $\pm $0.48 & 6.93 $\pm $0.26 & {} & {}   \\ 
        J1044-1826 & {} & {} & 4.4 $\pm $0.5 & 2.18 $\pm $0.14 & 2.64 $\pm $0.28 & 2.19 $\pm $0.24 & 3.14 $\pm $0.54 & 1594 $\pm $26 & 1527 $\pm $20 & {} & 6.76 $\pm $0.75 & 5.9 $\pm $0.4 & {} & {}   \\ 
        J1057-0805 & {} & {} & {} & {} & {} & {} & {} & {} & {} & {} & {} & {} & {} & {}   \\ 
        J1147-2145 & {} & {} & 5.7 $\pm $0.5 & 2.90 $\pm $0.13 & 3.28 $\pm $0.25 & 2.87 $\pm $0.16 & 3.21 $\pm $0.30 & 2151 $\pm $31 & 2044 $\pm $16 & {} & 8.56 $\pm $0.44 & 8.12 $\pm $0.25 & {} & {}   \\ 
        J2021-2235 & {} & {} & 24.6 $\pm $0.9 & 9.27 $\pm $0.16 & 9.74 $\pm $0.29 & 12.80 $\pm $0.15 & 14.42 $\pm $0.29 & 9469 $\pm $48 & 8924 $\pm $25 & {} & 41.53 $\pm $0.56 & 40.90 $\pm $0.32 & {} & {}   \\ 
        J2115-1417 & {} & {} & {} & {} & {} & {} & {} & {} & {} & {} & {} & {} & {} & {}   \\ 
        J2136-0116 & 3.34 & 4.76 & {} & {} & {} & {} & {} & {} & {} & {} & 9.37 $\pm $0.92 & 7.94 $\pm $0.47 & 47.8 $\pm $6.2 & 22.0 $\pm $3.6   \\ 
        J2137-1112 & {} & {} & {} & {} & {} & {} & {} & {} & {} & {} & {} & {} & {} & {}   \\ 
        J2143-2958 & {} & {} & {} & {} & {} & {} & {} & 212 $\pm $17 & 133.3 $\pm $6.6 & {} & {} & {} & {} & {}   \\ 
        J2155-1210 & {} & {} & {} & {} & {} & {} & {} & {} & {} & {} & {} & {} & {} & {}   \\ 
        J2207-2824 & {} & {} & {} & {} & {} & {} & {} & {} & {} & {} & {} & {} & {} & {}   \\ 
        J2229-1401 & {} & {} & {} & {} & {} & {} & {} & {} & {} & {} & {} & {} & {} & {}   \\ 
        J2244-1822 & {} & {} & {} & {} & {} & {} & {} & {} & {} & {} & {} & {} & {} & {}   \\ 
        J2250-1152 & {} & {} & {} & {} & {} & {} & {} & {} & {} & {} & {} & {} & {} & {}   \\ 
        J2311-2022 & {} & {} & {} & {} & {} & {} & {} & {} & {} & {} & {} & {} & {} & {}   \\ 
        J2358-1028 & {} & {} & {} & {} & {} & {} & {} & {} & {} & {} & {} & {} & {} & {}   \\ \hline
    \end{tabular}
    }
    \tablefoot{Columns: (1) Source name; (2) peak flux density in FIRST ($\mu$Jy beam$^{-1}$); (3) integrated flux density in FIRST ($\mu$Jy beam$^{-1}$); (4) peak flux density in NVSS ($\mu$Jy beam$^{-1}$); (5) peak flux density in VL1 ($\mu$Jy beam$^{-1}$); (6) integrated flux density in VL1 ($\mu$Jy beam$^{-1}$); (7) peak flux density in VL2 ($\mu$Jy beam$^{-1}$); (8) integrated flux density in VL2 ($\mu$Jy beam$^{-1}$); (9) integrated flux density in S5 ($\mu$Jy beam$^{-1}$); (10) peak flux density in S5 ($\mu$Jy beam$^{-1}$); (11) peak flux density in S9 ($\mu$Jy beam$^{-1}$); (12) integrated flux density in RACS ($\mu$Jy beam$^{-1}$); (13) peak flux density in RACS ($\mu$Jy beam$^{-1}$); (14) integrated flux density in TGSS ($\mu$Jy beam$^{-1}$); (15) peak flux density in TGSS ($\mu$Jy beam$^{-1}$).}
    \label{tab:other_data}
\end{sidewaystable*}

\clearpage

\begin{table*}
     \centering
    \footnotesize
    \caption{Luminosity of the sources with at least one detection at 15, 22, or 33 GHz.}
    \begin{tabular}{l c c c c c c c}
Name & MJD & $\log{L_p}$ (Ku) & $\log{L_i}$ (Ku) & $\log{L_p}$ (K) & $\log{L_i}$ (K) & $\log{L_p}$ (Ka) & $\log{L_i}$ (Ka)  \\
\hline\hline
J0000-0541 & 59904 & $39.11\pm0.03$ & $39.24\pm0.04$ & $38.97\pm0.04$ & $39.44\pm0.05$ & $39.07\pm0.04$ & $39.12\pm0.07$ \\
J0022-1039 & 59904 & $40.22\pm0.07$ & $40.56\pm0.12$ & $39.89\pm0.05$ & $40.24\pm0.07$ & $<40.21$ & {} \\
J0203-1247 & 59863 & $38.39\pm0.04$ & $38.43\pm0.05$ & $38.25\pm0.05$ & $38.28\pm0.08$ & $38.63\pm0.04$ & $38.77\pm0.07$ \\
J0203-1247 & 59902 & $38.29\pm0.03$ & $38.45\pm0.04$ & $38.39\pm0.04$ & $38.44\pm0.06$ & $38.54\pm0.03$ & $38.58\pm0.05$ \\
J0212-0201 & 59863 & $39.69\pm0.04$ & $39.92\pm0.06$ & $<40.20$ & {} & $<40.48$ & {} \\
J0212-0201 & 59902 & $39.73\pm0.05$ & $39.99\pm0.07$ & $<40.10$ & {} & $<40.33$ & {} \\
J0213-0551 & 59863 & $38.72\pm0.05$ & $38.87\pm0.09$ & $<39.23$ & {} & $<39.47$ & {} \\
J0213-0551 & 59902 & $38.81\pm0.05$ & $39.00\pm0.07$ & $<39.10$ & {} & $<39.33$ & {} \\
J0230-0859 & 59863 & $37.64\pm0.03$ & $37.70\pm0.03$ & $37.72\pm0.03$ & $37.88\pm0.04$ & $37.70\pm0.05$ & $37.86\pm0.07$ \\
J0230-0859 & 59902 & $37.61\pm0.02$ & $37.68\pm0.03$ & $37.56\pm0.03$ & $37.76\pm0.04$ & $37.67\pm0.04$ & $37.74\pm0.06$ \\
J0239-1118 & 59863 & $39.71\pm0.03$ & $39.79\pm0.05$ & $40.04\pm0.04$ & $40.03\pm0.06$ & $40.30\pm0.04$ & $40.32\pm0.05$ \\
J0239-1118 & 59902 & $39.75\pm0.03$ & $39.65\pm0.04$ & $39.93\pm0.04$ & $40.05\pm0.06$ & $40.22\pm0.04$ & $40.28\pm0.06$ \\
J0400-2500 & 59872 & $39.18\pm0.03$ & $39.24\pm0.04$ & $39.09\pm0.04$ & $39.31\pm0.06$ & $<39.24$ & {} \\
J0400-2500 & 59882 & $39.12\pm0.03$ & $39.23\pm0.04$ & $38.94\pm0.05$ & $39.20\pm0.06$ & $<39.17$ & {} \\
J0413-0050 & 59872 & $<37.61$ & {} & $<37.94$ & {} & $<38.24$ & {} \\
J0413-0050 & 59882 & $37.70\pm0.06$ & $37.79\pm0.10$ & * & {} & $<38.24$ & {} \\
J0422-1854 & 59872 & $38.72\pm0.03$ & $38.81\pm0.03$ & $38.72\pm0.04$ & $38.91\pm0.07$ & $38.82\pm0.04$ & $38.79\pm0.08$ \\
J0422-1854 & 59882 & $38.74\pm0.03$ & $38.81\pm0.04$ & $38.80\pm0.04$ & $38.82\pm0.06$ & $<38.78$ & {} \\
J0436-1022 & 59872 & $38.86\pm0.02$ & $38.90\pm0.02$ & $38.86\pm0.02$ & $38.92\pm0.02$ & $38.91\pm0.03$ & $38.99\pm0.04$ \\
J0436-1022 & 59882 & $38.85\pm0.02$ & $38.89\pm0.02$ & $38.88\pm0.02$ & $38.92\pm0.03$ & $38.98\pm0.03$ & $39.03\pm0.03$ \\
J0447-0508 & 59872 & $38.95\pm0.02$ & $39.04\pm0.03$ & $38.91\pm0.03$ & $39.04\pm0.03$ & $38.93\pm0.04$ & $39.08\pm0.05$ \\
J0447-0508 & 59882 & $38.98\pm0.02$ & $39.03\pm0.03$ & $38.99\pm0.03$ & $39.10\pm0.03$ & $38.94\pm0.03$ & $39.01\pm0.05$ \\
J0452-2953 & 59872 & $40.58\pm0.06$ & $40.79\pm0.04$ & $40.19\pm0.11$ & $40.73\pm0.07$ & $40.26\pm0.11$ & $40.64\pm0.09$ \\
J0452-2953 & 59882 & $40.44\pm0.06$ & $40.74\pm0.08$ & $40.33\pm0.09$ & $40.70\pm0.14$ & $40.36\pm0.08$ & $40.59\pm0.12$ \\
J0549-2425 & 59898 & $38.62\pm0.02$ & $38.66\pm0.03$ & $38.66\pm0.03$ & $38.70\pm0.04$ & $38.75\pm0.03$ & $38.77\pm0.05$ \\
J0549-2425 & 59899 & $38.63\pm0.02$ & $38.69\pm0.03$ & $38.63\pm0.03$ & $38.70\pm0.05$ & $38.71\pm0.03$ & $38.79\pm0.05$ \\
J0622-2317 & 59898 & $37.57\pm0.06$ & $37.95\pm0.06$ & $<37.82$ & {} & $<38.24$ & {} \\
J0622-2317 & 59899 & $<37.61$ & {} & $<37.82$ & {} & $<38.24$ & {} \\
J0820-1741 & 59883 & $38.36\pm0.05$ & $38.62\pm0.08$ & $<38.52$ & {} & $<38.87$ & {} \\
J0820-1741 & 59904 & $38.50\pm0.04$ & $38.61\pm0.06$ & $38.55\pm0.05$ & $38.76\pm0.09$ & $<38.86$ & {} \\
J0842-0349 & 59883 & $38.59\pm0.05$ & $39.01\pm0.07$ & $<38.99$ & {} & $<39.16$ & {} \\
J0842-0349 & 59904 & $38.56\pm0.06$ & $38.88\pm0.08$ & $<38.88$ & {} & $<39.16$ & {} \\
J0846-1214 & 59883 & $39.96\pm0.03$ & $39.96\pm0.03$ & $39.97\pm0.03$ & $39.98\pm0.03$ & $39.90\pm0.03$ & $40.00\pm0.04$ \\
J0846-1214 & 59904 & $39.97\pm0.02$ & $39.97\pm0.02$ & $39.96\pm0.02$ & $39.99\pm0.03$ & $39.94\pm0.03$ & $39.98\pm0.03$ \\
J0849-2351 & 59883 & $39.39\pm0.03$ & $39.42\pm0.05$ & $39.10\pm0.06$ & $39.41\pm0.09$ & $<39.38$ & {} \\
J0849-2351 & 59904 & $39.38\pm0.03$ & $39.49\pm0.04$ & $39.19\pm0.04$ & $39.30\pm0.07$ & $39.39\pm0.05$ & $39.56\pm0.08$ \\
J0850-0318 & 59883 & $39.21\pm0.03$ & $39.27\pm0.05$ & $<39.26$ & {} & $<39.58$ & {} \\
J0850-0318 & 59904 & $39.25\pm0.04$ & $39.46\pm0.05$ & $<39.15$ & {} & $<39.50$ & {} \\
J1032-2707 & 59862 & $38.94\pm0.04$ & $39.32\pm0.04$ & $38.79\pm0.05$ & $39.11\pm0.06$ & $<38.96$ & {} \\
J1032-2707 & 59903 & $39.04\pm0.03$ & $39.28\pm0.03$ & $38.96\pm0.03$ & $39.30\pm0.04$ & $38.99\pm0.04$ & $39.05\pm0.07$ \\
J1044-1826 & 59862 & $39.55\pm0.03$ & $39.52\pm0.03$ & $39.52\pm0.03$ & $39.53\pm0.05$ & $39.52\pm0.05$ & $39.62\pm0.09$ \\
J1044-1826 & 59903 & $39.51\pm0.02$ & $39.51\pm0.03$ & $39.47\pm0.03$ & $39.47\pm0.05$ & $39.48\pm0.04$ & $39.37\pm0.07$ \\
J1147-2145 & 59903 & $40.19\pm0.03$ & $40.20\pm0.04$ & $40.15\pm0.04$ & $40.12\pm0.06$ & $40.12\pm0.05$ & $40.16\pm0.08$ \\
J2021-2235 & 59900 & $40.62\pm0.03$ & $40.62\pm0.03$ & $40.53\pm0.03$ & $40.59\pm0.04$ & $40.50\pm0.03$ & $40.57\pm0.05$ \\
J2244-1822 & 59884 & $39.19\pm0.06$ & $39.43\pm0.09$ & $<39.46$ & {} & $<39.71$ & {} \\
J2358-1029 & 59884 & $38.91\pm0.05$ & $38.98\pm0.10$ & $<39.31$ & {} & $<39.56$ & {} \\
\hline
   \end{tabular}
    \tablefoot{Columns: (1) Source name; (2) modified Julian date (MJD) of the observation; (3) logarithm of the peak luminosity {at 15 GHz} (\ergs); (4) logarithm of the integrated luminosity {at 15 GHz} (\ergs); (5) logarithm of the peak luminosity {at 15 GHz} (\ergs); (6) logarithm of the integrated luminosity {at 15 GHz} (\ergs); (7) logarithm of the peak luminosity {at 15 GHz} (\ergs); (8) logarithm of the integrated luminosity {at 15 GHz} (\ergs). The asterisk indicates a technical problem {at 22 GHz} observation for J0413-0050, no data were taken in this band. }
    \label{tab:luminosity}
\end{table*}

\clearpage
\section{Radio maps}
In the following section, we include the radio maps of the sources that show some sign of extended emission, and the overlap between the {15 GHz} image and the Pan-STARRS i-band image. For J0452-2953, the map {at 15 GHz} observed on MJD 59872 is shown in Fig.~\ref{fig:J0452_map}, while its host galaxy in Fig.~\ref{fig:J0452_host}. For J1032-2707, the map {at 15 GHz} observed on MJD 59862 is shown in Fig.~\ref{fig:J1032_map}.

\begin{figure}[!h]
    \centering
    \includegraphics[width=\columnwidth, trim={0 11cm 0 1cm},clip]{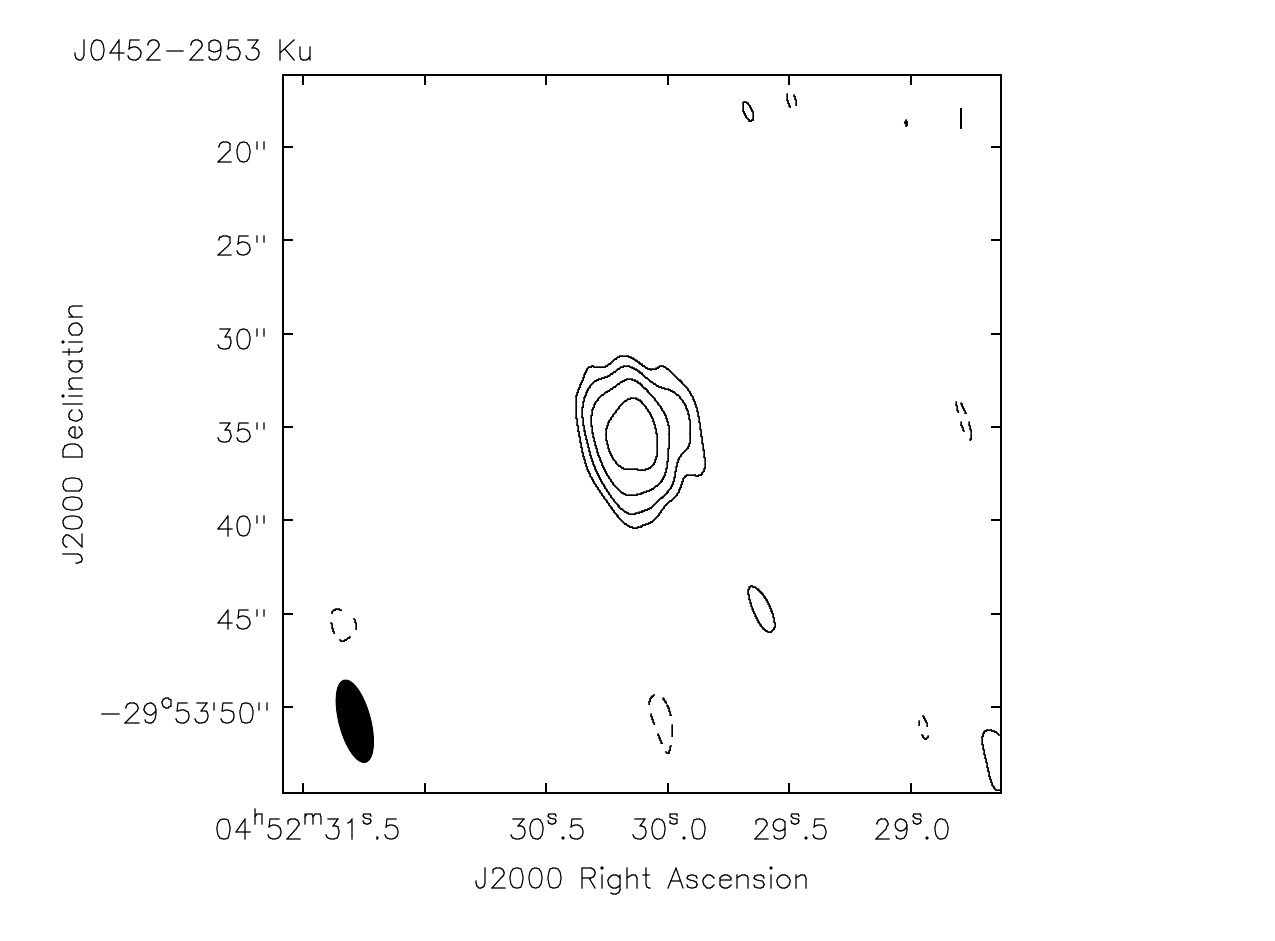}
    \caption{Radio map of J0452-2953 {at 15 GHz}, observed on MJD 59882. The map rms is $\sigma = 17 \mu$Jy, the contours are at [-3, 3, 6, 12, 24]$\times\sigma$.}
    \label{fig:J0452_map_K_59882}
\end{figure}

\begin{figure}[!h]
    \centering
    \includegraphics[width=\columnwidth, trim={0 11cm 0 1cm},clip]{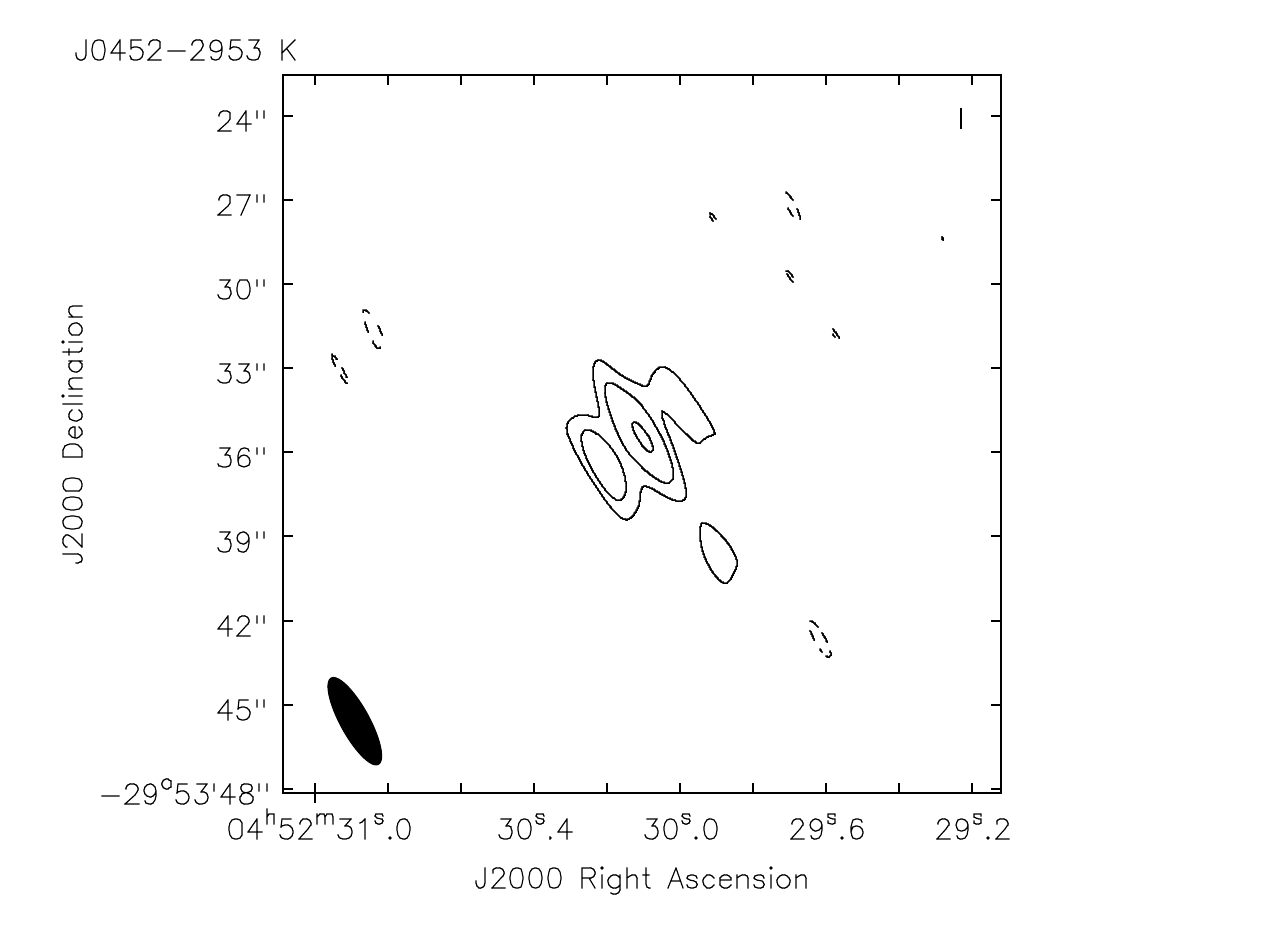}
    \caption{Radio map of J0452-2953 {at 22 GHz}, observed on MJD 59872. The map rms is $\sigma = 30 \mu$Jy, the contours are at [-3, 3, 6, 12]$\times\sigma$.}
    \label{fig:J0452_map_K_59872}
\end{figure}

\begin{figure}[!h]
    \centering
    \includegraphics[width=\columnwidth, trim={0 11cm 0 1cm},clip]{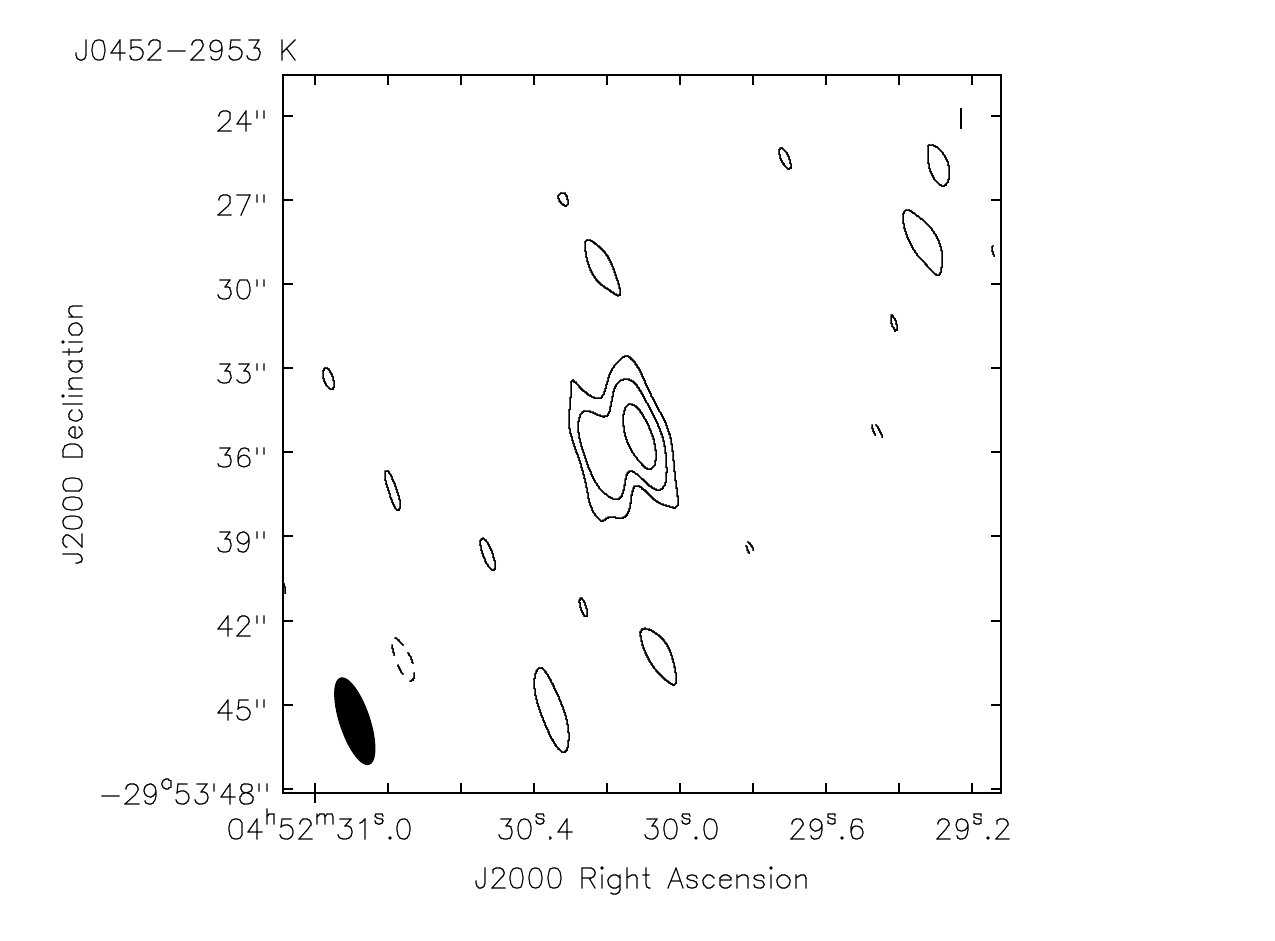}
    \caption{Radio map of J0452-2953 {at 22 GHz}, observed on MJD 59882. The map rms is $\sigma = 26 \mu$Jy, the contours are at [-3, 3, 6, 12]$\times\sigma$.}
    \label{fig:J0452_map_K_59882}
\end{figure}

\begin{figure}[!h]
    \centering
    \includegraphics[width=\columnwidth, trim={0 11cm 0 1cm},clip]{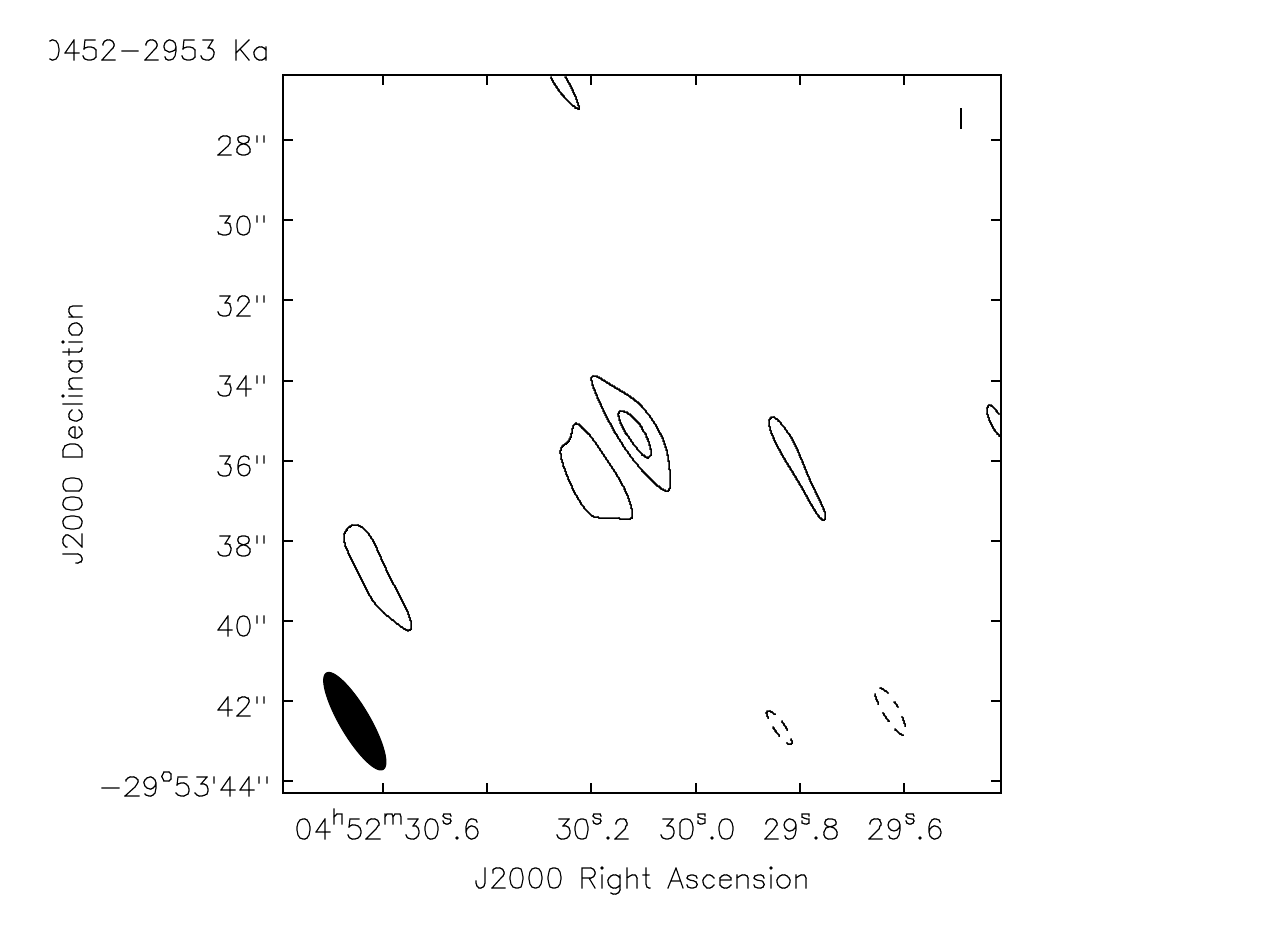}
    \caption{Radio map of J0452-2953 {at 33 GHz}. The map rms is $\sigma = 39 \mu$Jy, the contours are at [-3, 3, 6]$\times\sigma$.}
    \label{fig:J0452_map_Ka_59882}
\end{figure}

\begin{figure}[!h]
    \centering
    \includegraphics[width=\columnwidth, trim={0 11cm 0 1cm},clip]{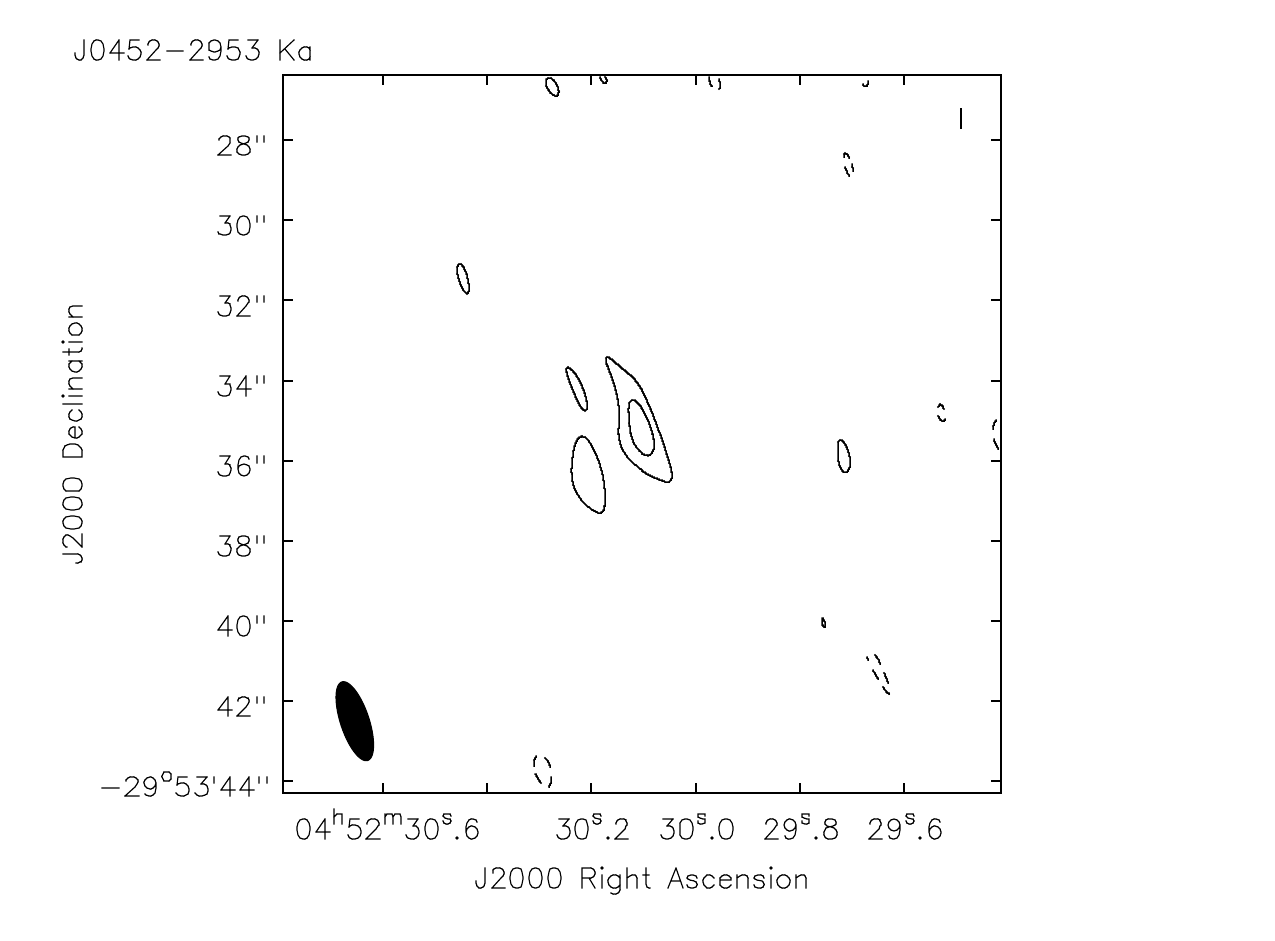}
    \caption{Radio map of J0452-2953 {at 33 GHz}. The map rms is $\sigma = 36 \mu$Jy, the contours are at [-3, 3, 6]$\times\sigma$.}
    \label{fig:J0452_map_Ka_59882}
\end{figure}

\begin{figure}[!h]
    \centering
    \includegraphics[width=\columnwidth, trim={0 11cm 0 1cm},clip]{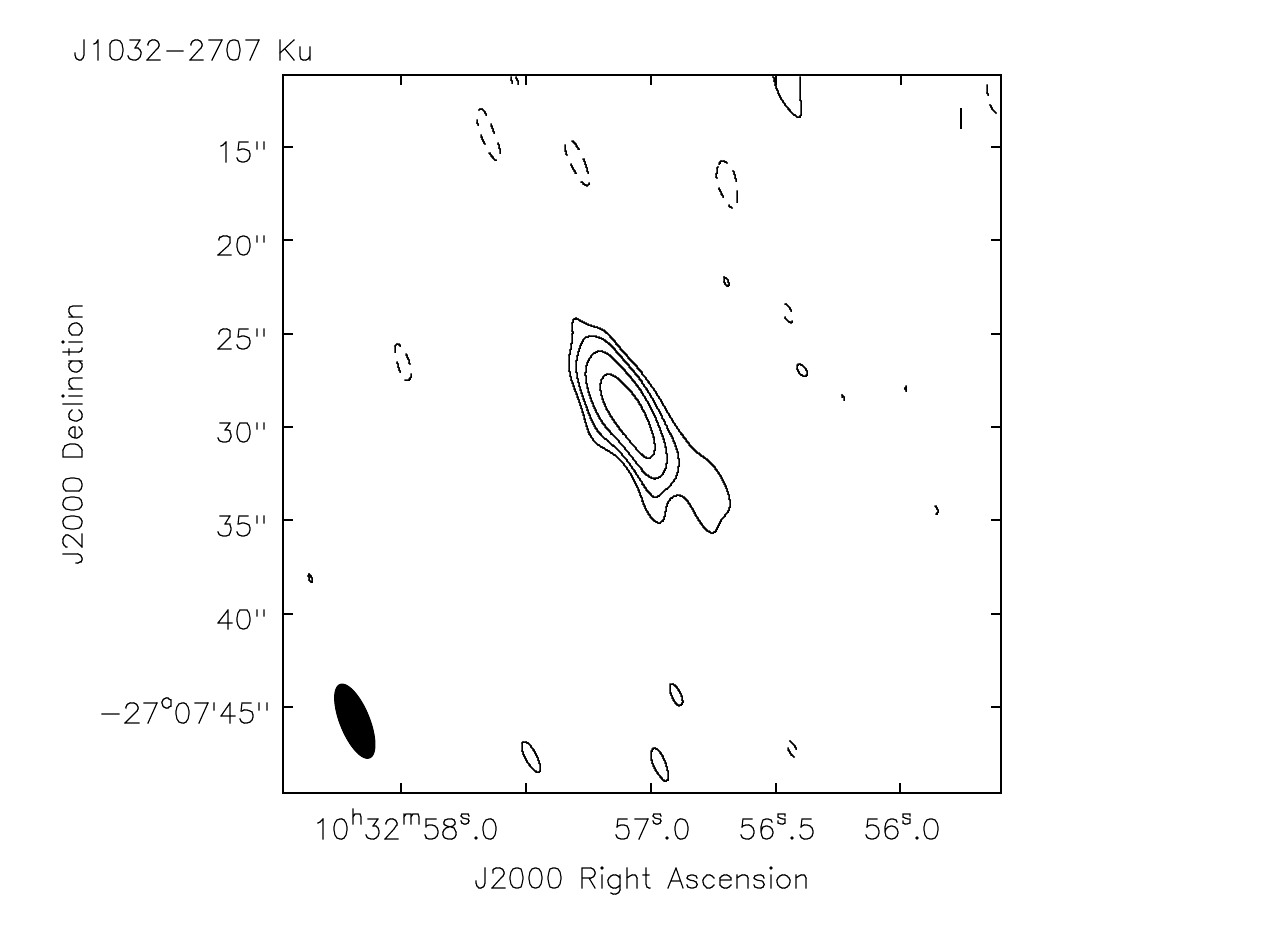}
    \caption{Radio map of J1032-2707 {at 15 GHz}, observed on MJD 59903. The map rms is $\sigma = 15 \mu$Jy, the contours are at [-3, 3, 6, 12, 24]$\times\sigma$.}
    \label{fig:J1032_map_Ku_59903}
\end{figure}

\begin{figure}[!h]
    \centering
    \includegraphics[width=\columnwidth, trim={0 11cm 0 1cm},clip]{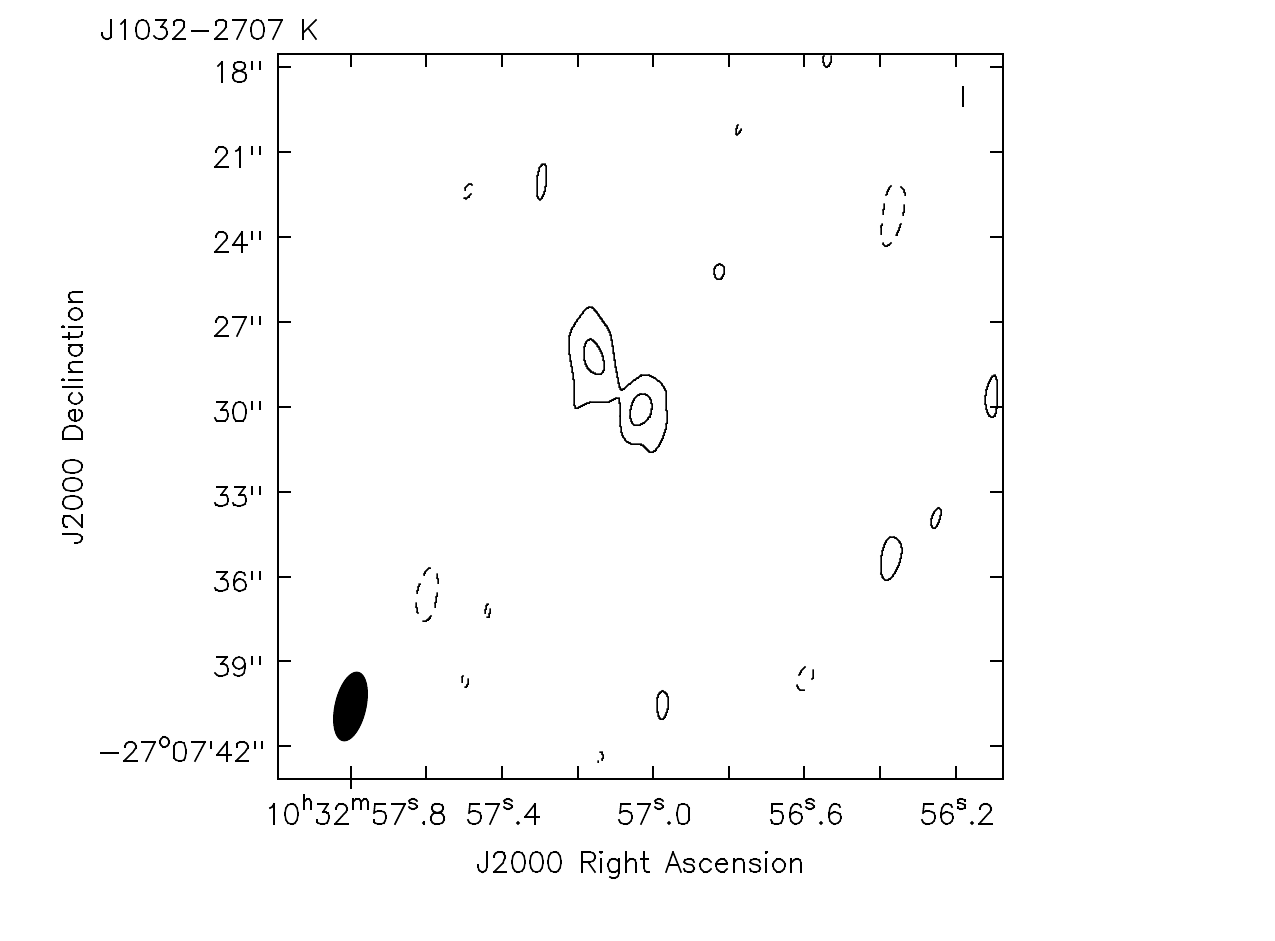}
    \caption{Radio map of J1032-2707 {at 22 GHz}, observed on MJD 59862. The map rms is $\sigma = 35 \mu$Jy, the contours are at [-3, 3, 6]$\times\sigma$.}
    \label{fig:J1032_map_K_59862}
\end{figure}

\begin{figure}[!h]
    \centering
    \includegraphics[width=\columnwidth, trim={0 11cm 0 1cm},clip]{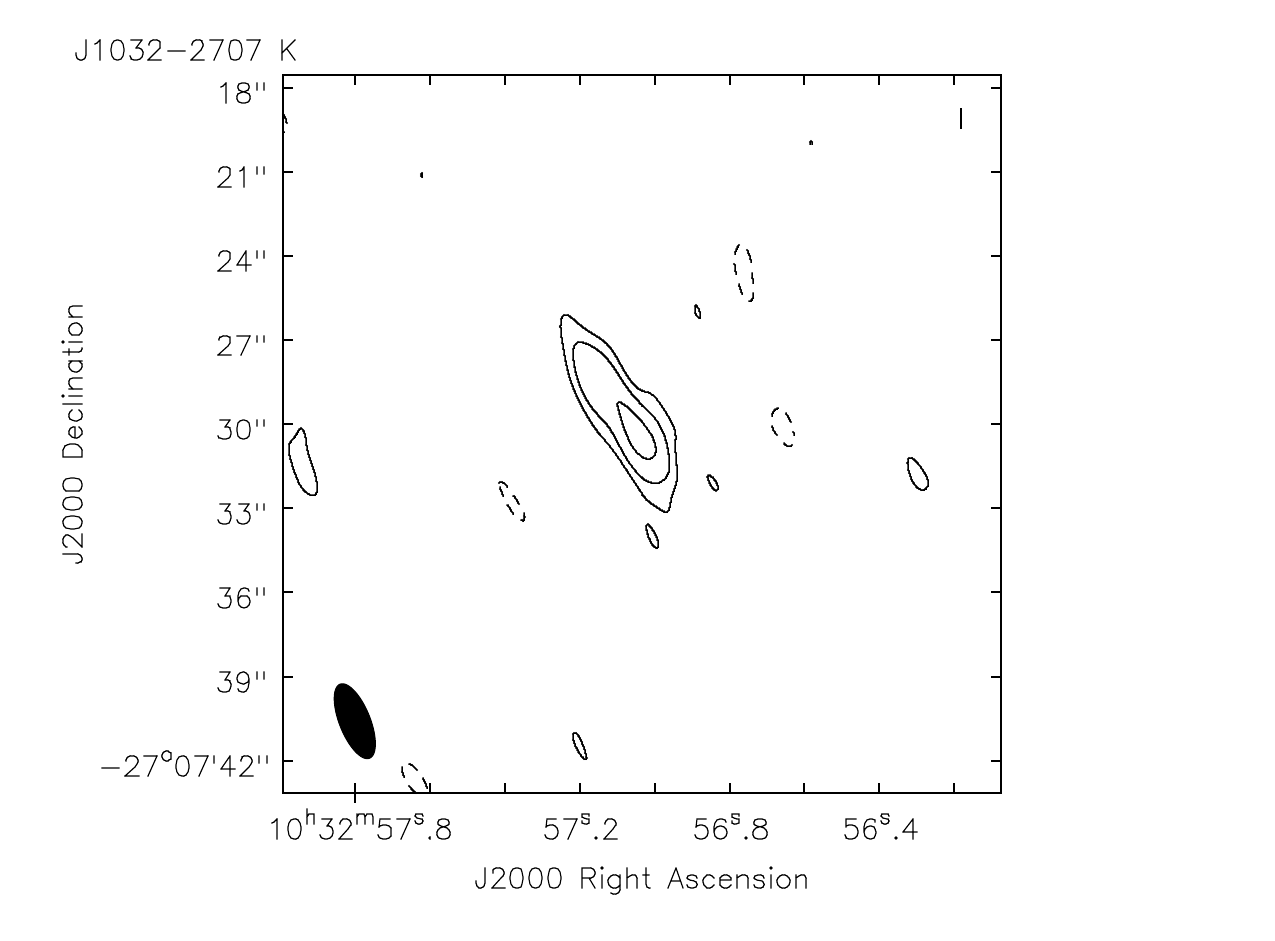}
    \caption{Radio map of J1032-2707 {at 22 GHz}, observed on MJD 59903. The map rms is $\sigma = 22 \mu$Jy, the contours are at [-3, 3, 6, 12]$\times\sigma$.}
    \label{fig:J1032_map_K_59903}
\end{figure}

\begin{figure}[!h]
    \centering
    \includegraphics[width=\columnwidth, trim={0 11cm 0 1cm},clip]{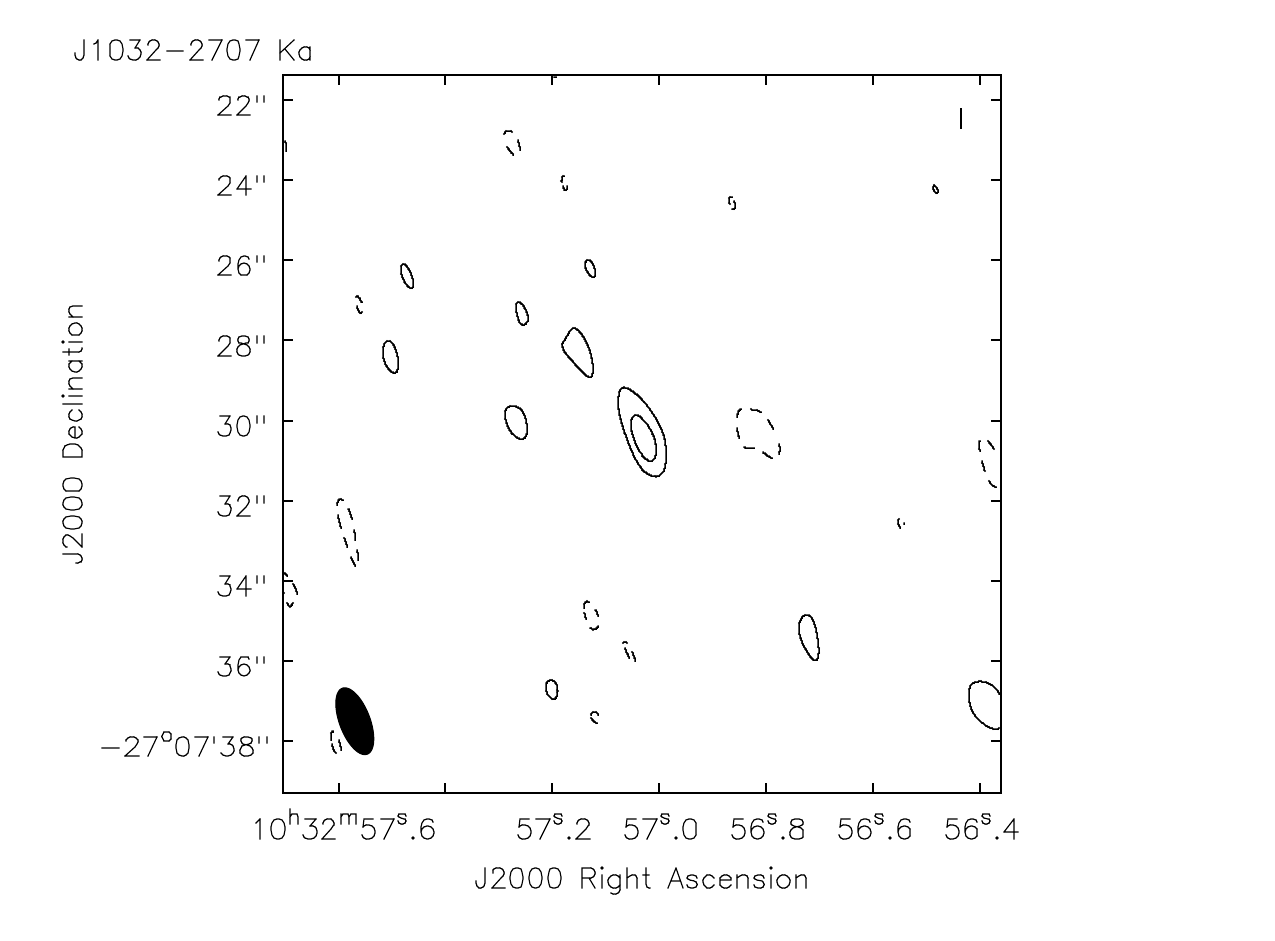}
    \caption{Radio map of J1032-2707 {at 33 GHz}, observed on MJD 59903. The map rms is $\sigma = 30 \mu$Jy, the contours are at [-3, 3, 6]$\times\sigma$.}
    \label{fig:J1032_map_Ka_59903}
\end{figure}

\begin{figure}
    \centering
    \includegraphics[width=\columnwidth]{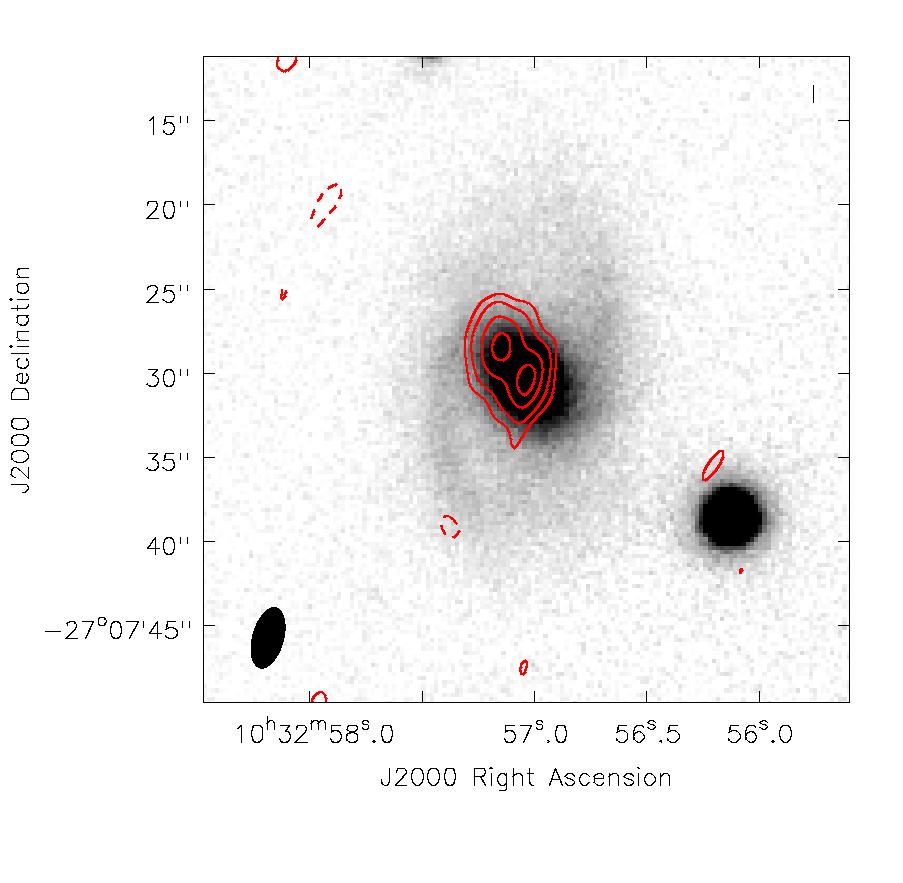}
    \caption{In red the contours of J1032-2707 as in Fig.~\ref{fig:J1032_map}, overlapped with the i-band image of its host galaxy extracted from Pan-STARRS.}
    \label{fig:J1032_host}
\end{figure}

\clearpage
\section{Radio spectra}
All the radio spectra of the sources detected in at least one {time at 15, 22, or 33 GHz} with {the} JVLA. The symbols are as follows. The empty points indicate a flux density (left y axis), while solid points indicate integrated fluxes (right y axis). The red solid line shows the fit with a power law, while the blue dashed line is the fit with a log parabola. The empty triangles represent upper limits, calculated as three times the level of the noise in the images, {and they are reported only for our new JVLA observations.}

\begin{figure}[!h]
    \centering
    \includegraphics[width=\columnwidth]{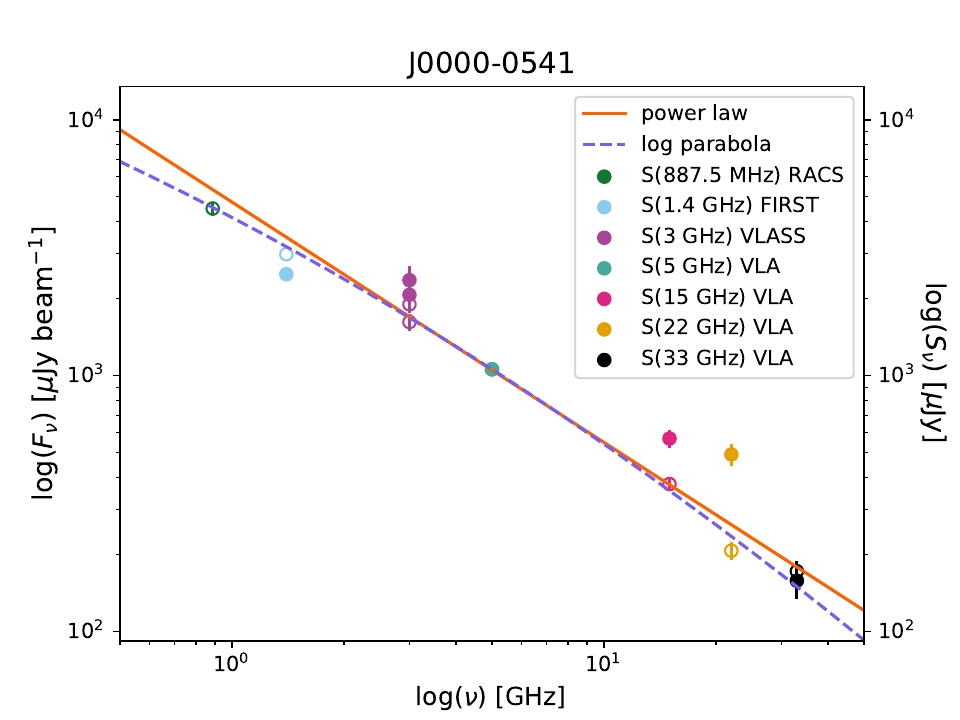}
    \caption{Spectrum of J0000-0541. Symbols as described in the text.}
    \label{fig:J0000-0541}
\end{figure}

\begin{figure}[!h]
    \centering
    \includegraphics[width=\columnwidth]{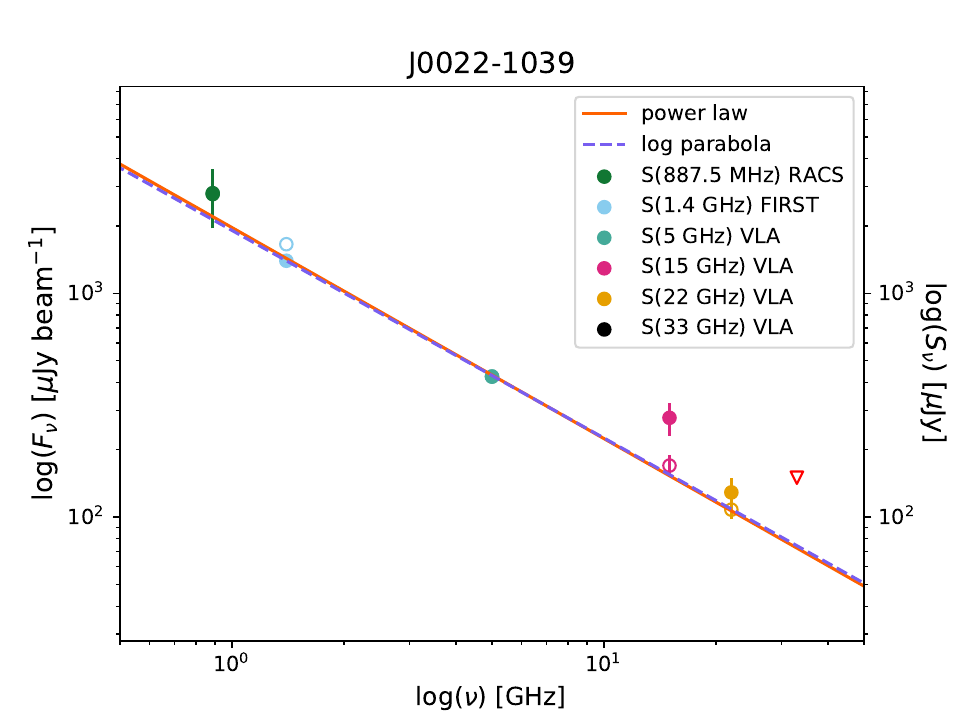}
    \caption{Spectrum of J0022-1039. Symbols as described in the text. }
    \label{fig:J0022-1039}
\end{figure}

\begin{figure}
    \centering
    \includegraphics[width=\columnwidth]{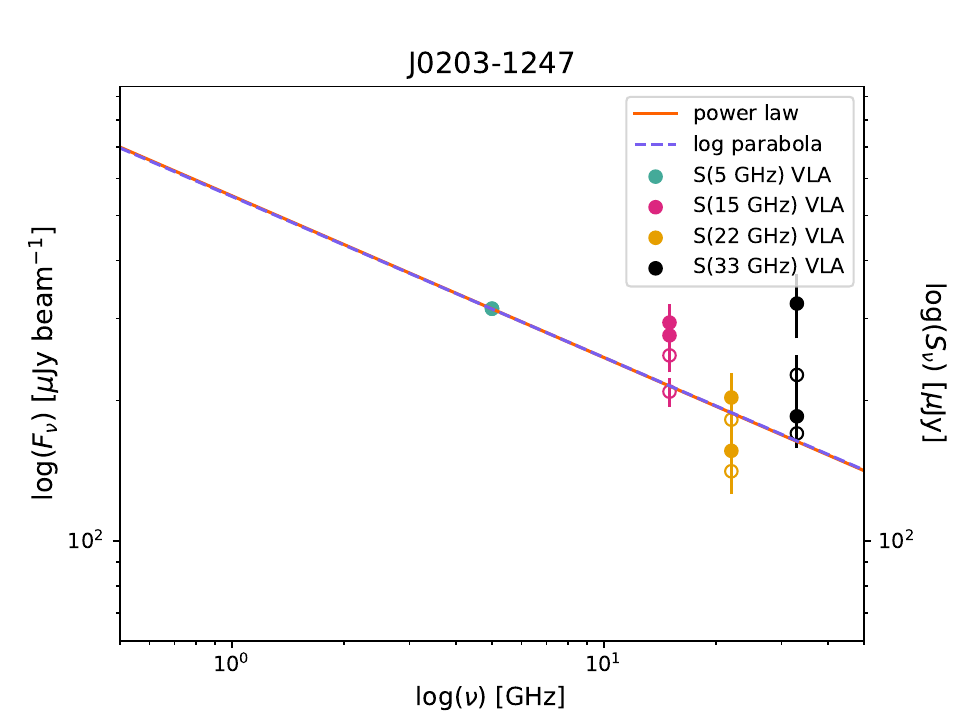}
    \caption{Spectrum of J0203-1247. Symbols as described in the text.}
    \label{fig:J0203-1247}
\end{figure}

\begin{figure}
    \centering
    \includegraphics[width=\columnwidth]{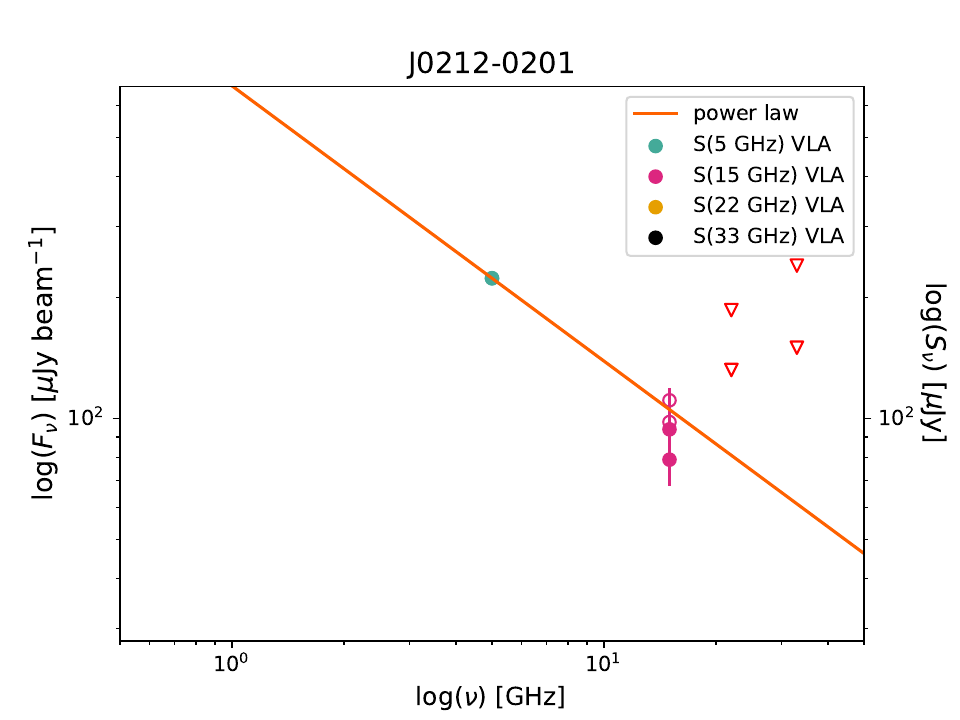}
    \caption{Spectrum of J0212-0201. Symbols as described in the text.}
    \label{fig:J0212-0201}
\end{figure}

\begin{figure}
    \centering
    \includegraphics[width=\columnwidth]{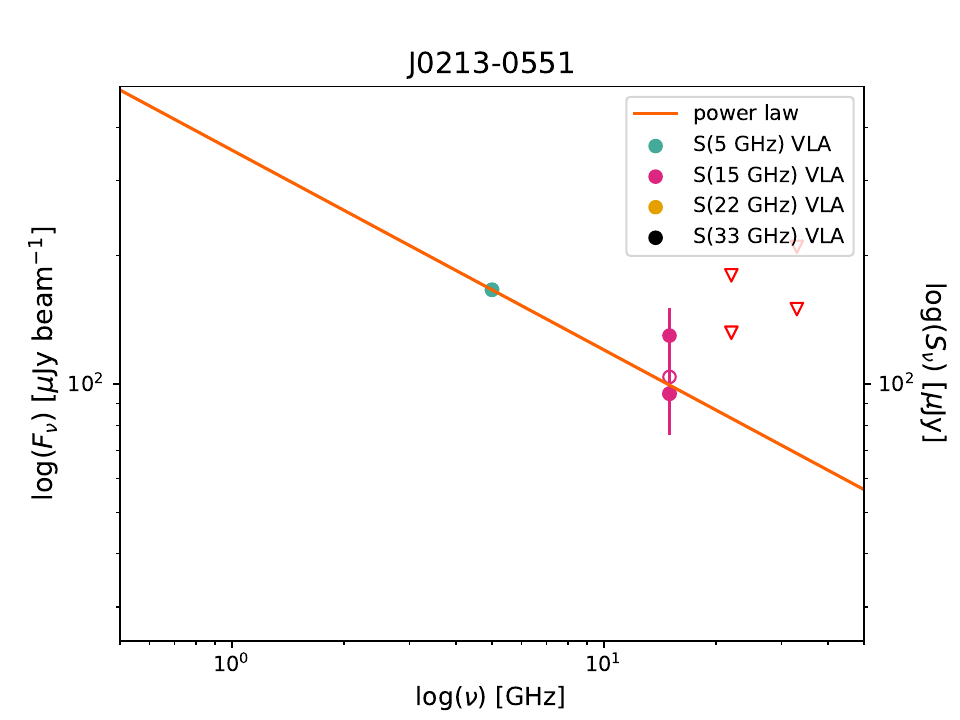}
    \caption{Spectrum of J0213-0551. Symbols as described in the text.}
    \label{fig:J0213-0551}
\end{figure}

\begin{figure}
    \centering
    \includegraphics[width=\columnwidth]{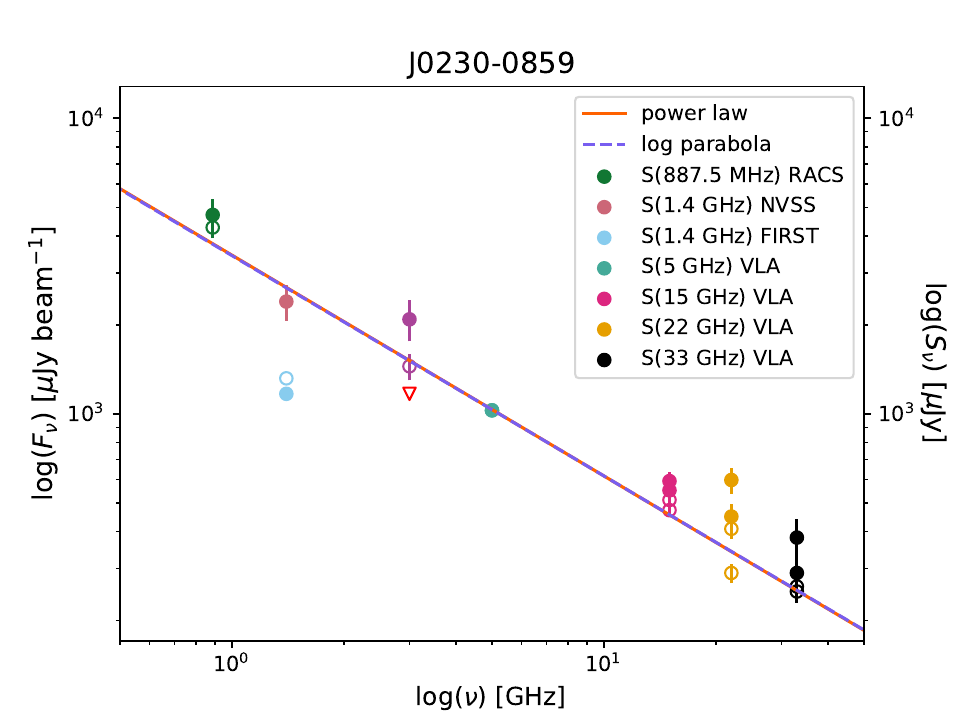}
    \caption{Spectrum of J0230-0859. Symbols as described in the text.}
    \label{fig:J0230-0859}
\end{figure}

\begin{figure}
    \centering
    \includegraphics[width=\columnwidth]{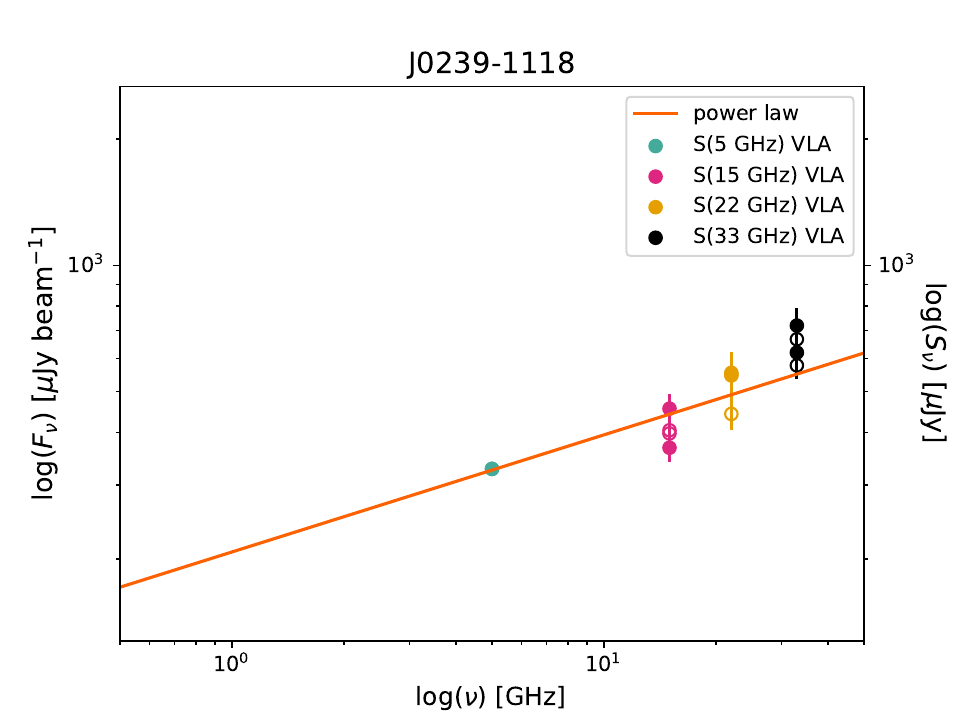}
    \caption{Spectrum of J0239-1118. Symbols as described in the text.}
    \label{fig:J0239-1118}
\end{figure}

\begin{figure}
    \centering
    \includegraphics[width=\columnwidth]{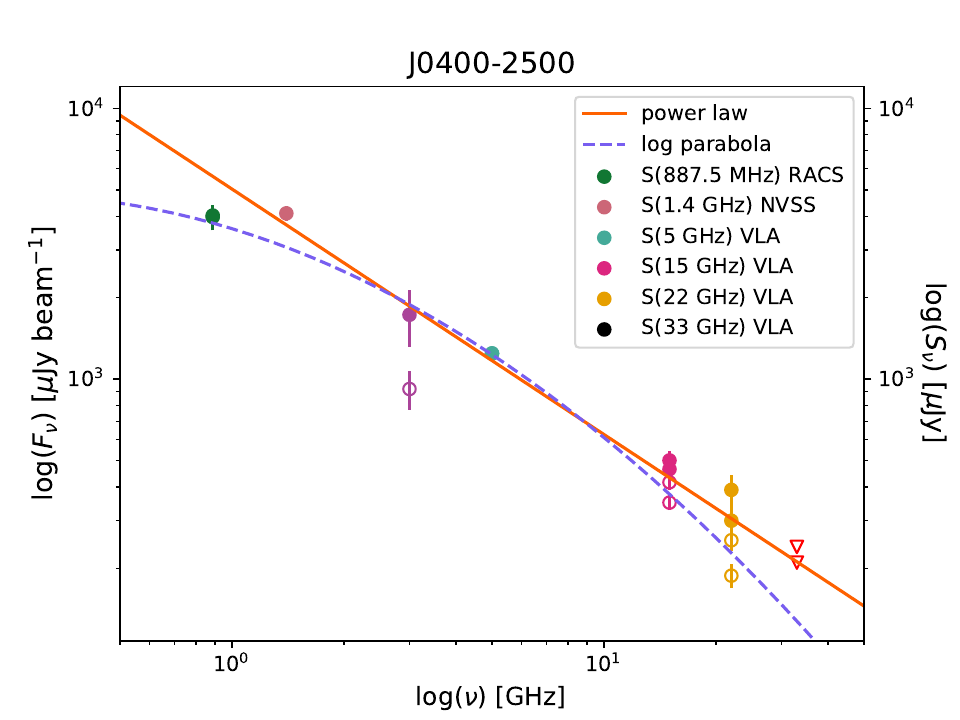}
    \caption{Spectrum of J0400-2500. Symbols as described in the text.}
    \label{fig:J0400-2500}
\end{figure}

\begin{figure}
    \centering
    \includegraphics[width=\columnwidth]{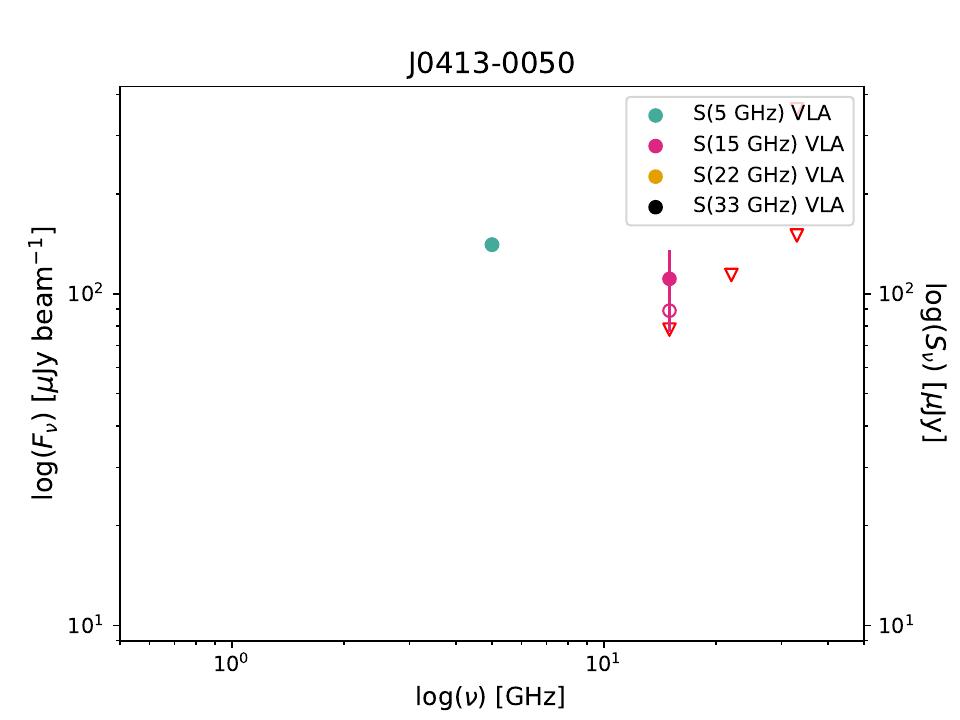}
    \caption{Spectrum of J0413-0050. Symbols as described in the text.}
    \label{fig:J0413-0050}
\end{figure}

\begin{figure}
    \centering
    \includegraphics[width=\columnwidth]{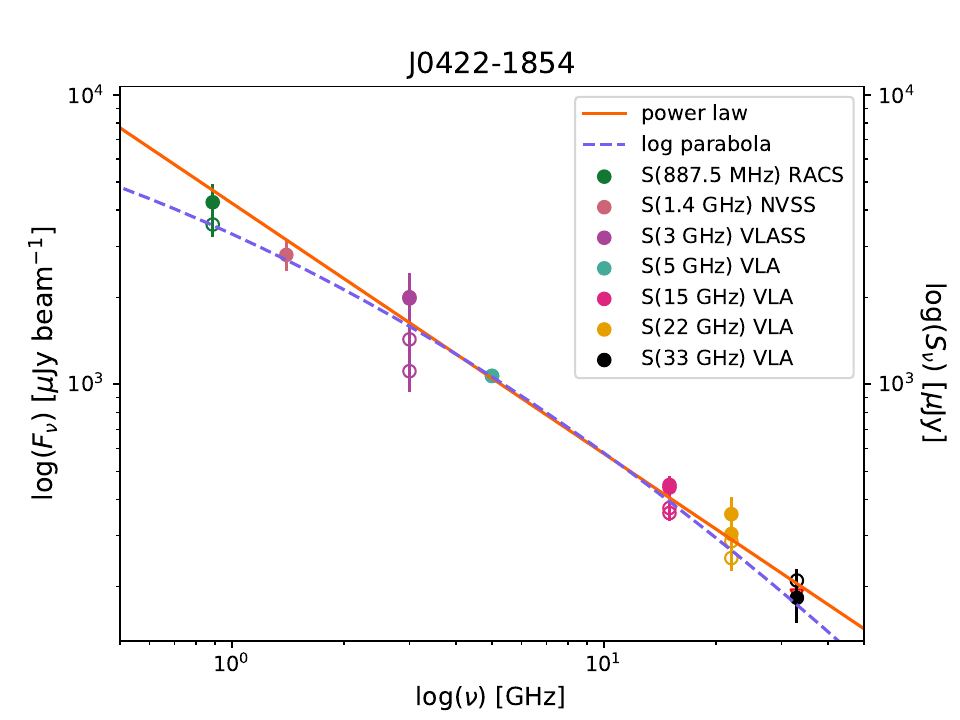}
    \caption{Spectrum of J0422-1854. Symbols as described in the text.}
    \label{fig:J0422-1854}
\end{figure}

\begin{figure}
    \centering
    \includegraphics[width=\columnwidth]{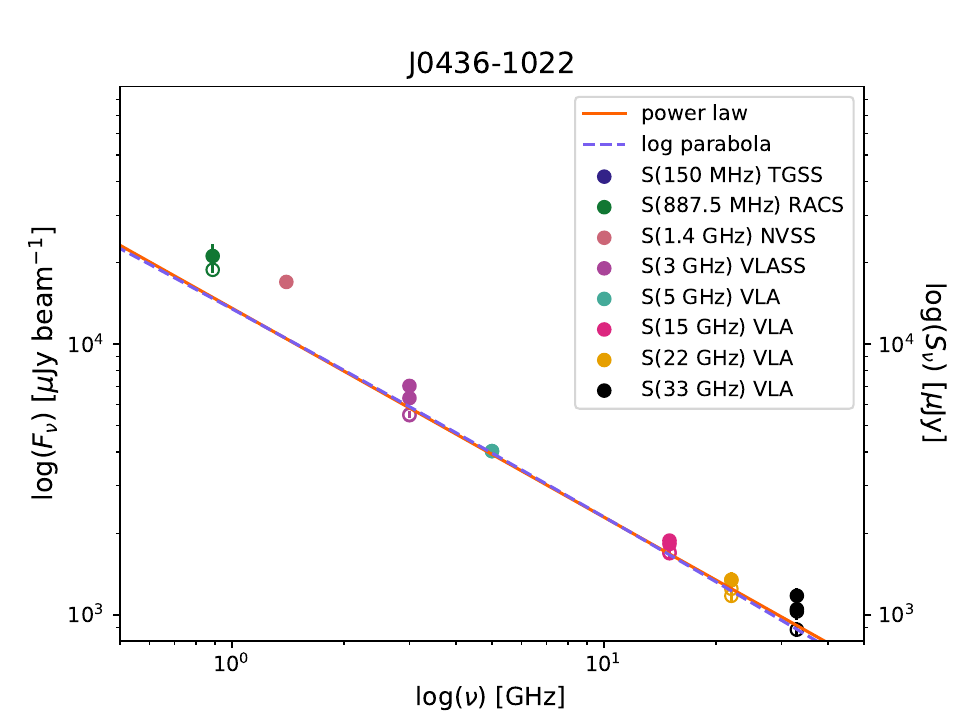}
    \caption{Spectrum of J0436-1022. Symbols as described in the text.}
    \label{fig:J0436-1022}
\end{figure}

\begin{figure}
    \centering
    \includegraphics[width=\columnwidth]{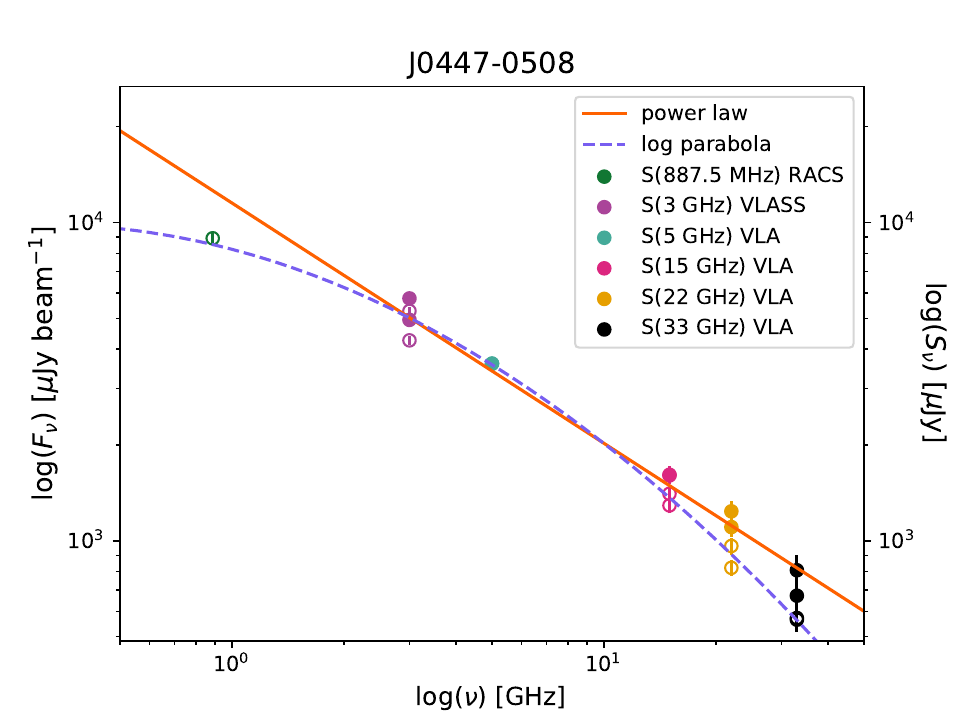}
    \caption{Spectrum of J0447-0508. Symbols as described in the text.}
    \label{fig:J0447-0508}
\end{figure}

\begin{figure}
    \centering
    \includegraphics[width=\columnwidth]{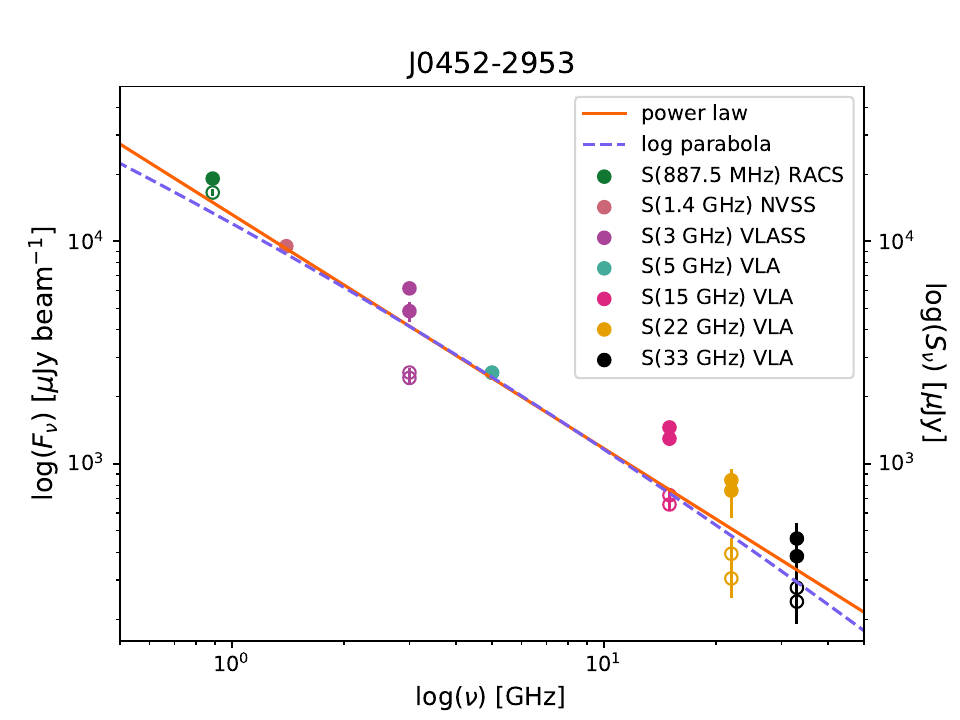}
    \caption{Spectrum of J0452-2953. Symbols as described in the text.}
    \label{fig:J0452-2953}
\end{figure}

\begin{figure}
    \centering
    \includegraphics[width=\columnwidth]{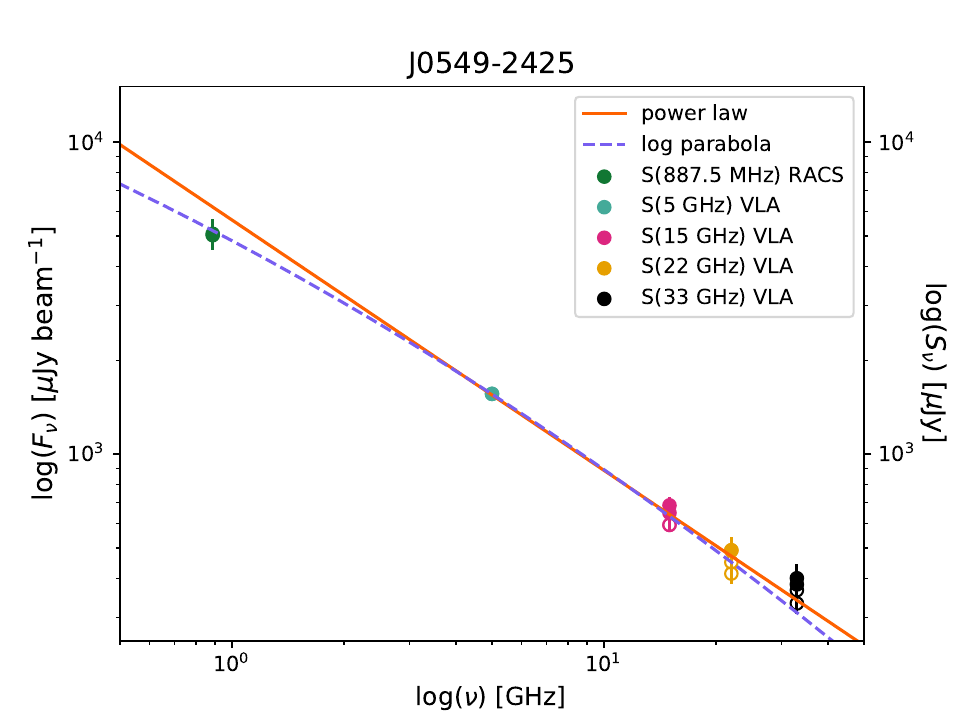}
    \caption{Spectrum of J0549-2425. Symbols as described in the text.}
    \label{fig:J0549-2425}
\end{figure}

\begin{figure}
    \centering
    \includegraphics[width=\columnwidth]{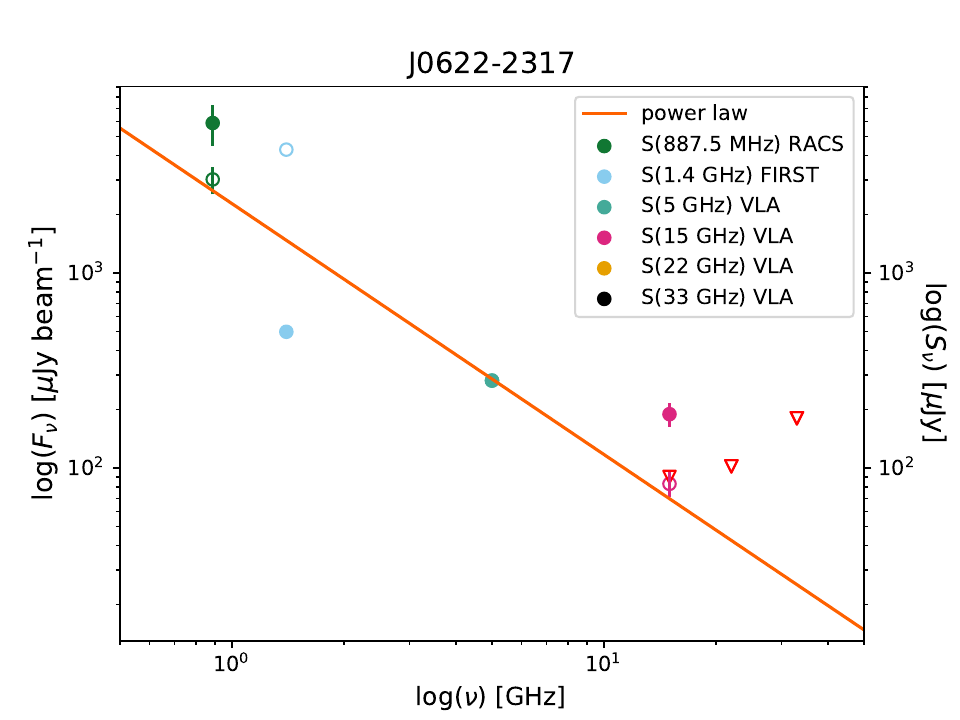}
    \caption{Spectrum of J0622-2317. Symbols as described in the text.}
    \label{fig:J0622-2317}
\end{figure}

\begin{figure}
    \centering
    \includegraphics[width=\columnwidth]{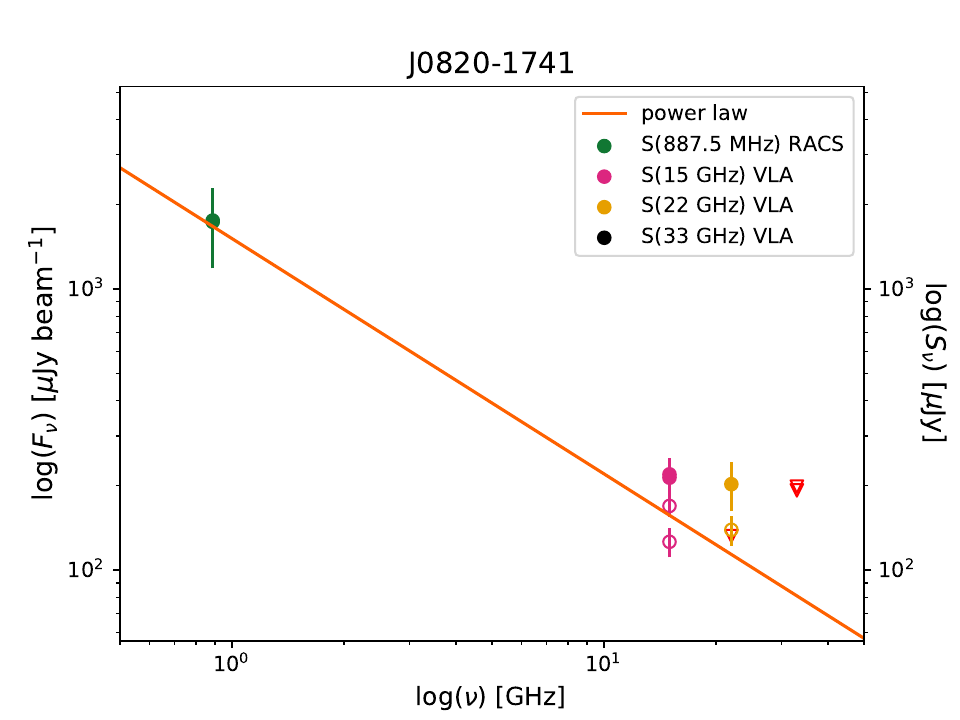}
    \caption{Spectrum of J0820-1741. Symbols as described in the text.}
    \label{fig:J0820-1741}
\end{figure}

\begin{figure}
    \centering
    \includegraphics[width=\columnwidth]{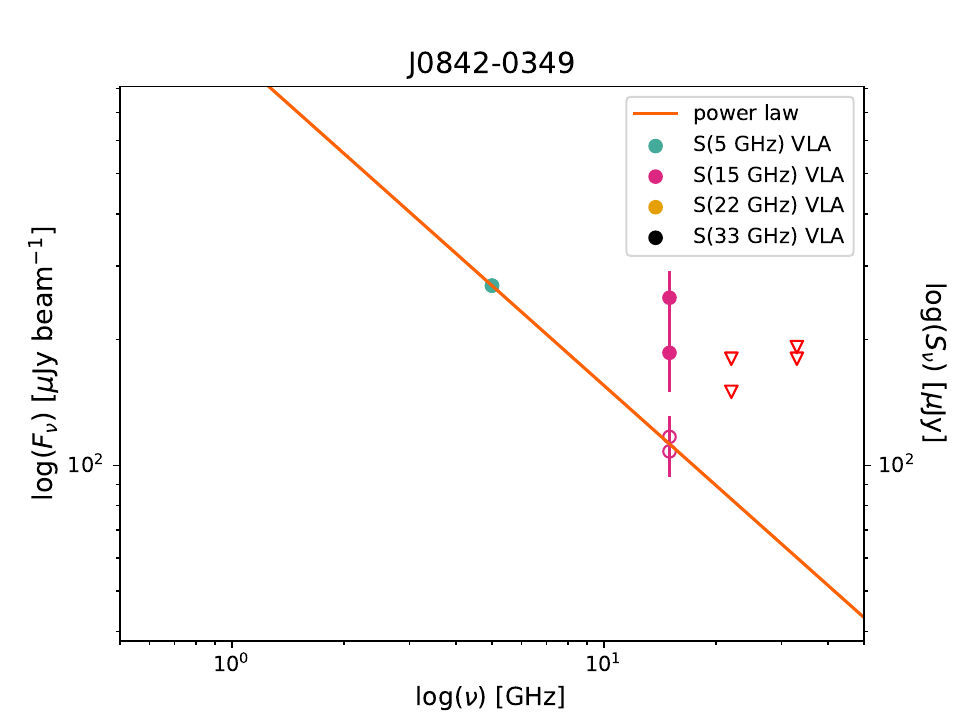}
    \caption{Spectrum of J0842-0349. Symbols as described in the text.}
    \label{fig:J0842-0349}
\end{figure}

\begin{figure}
    \centering
    \includegraphics[width=\columnwidth]{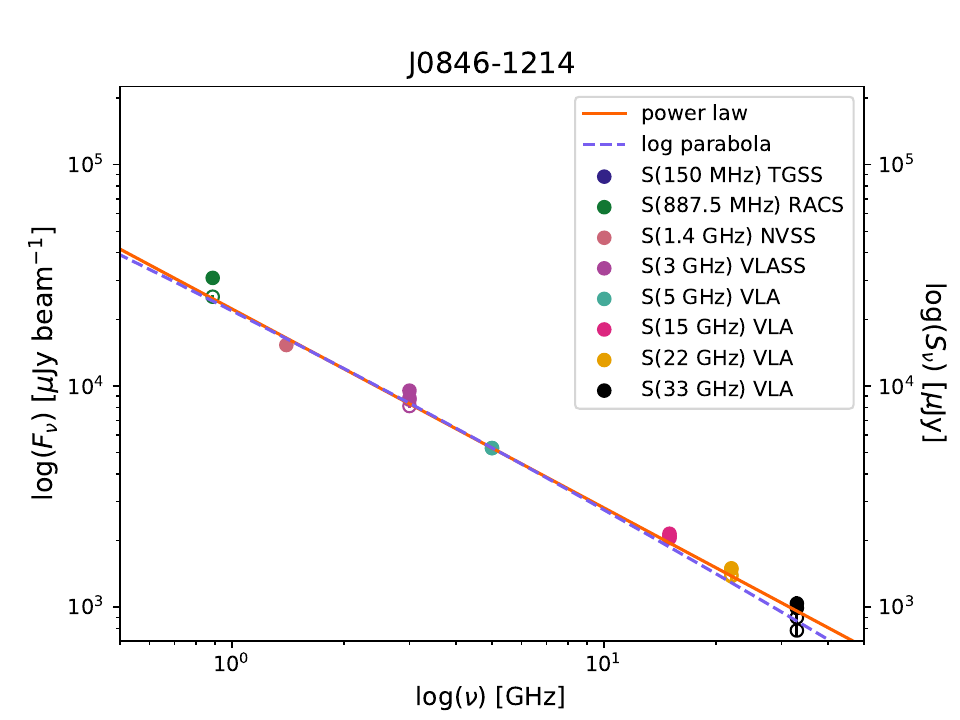}
    \caption{Spectrum of J0846-1214. Symbols as described in the text.}
    \label{fig:J0846-1214}
\end{figure}

\begin{figure}
    \centering
    \includegraphics[width=\columnwidth]{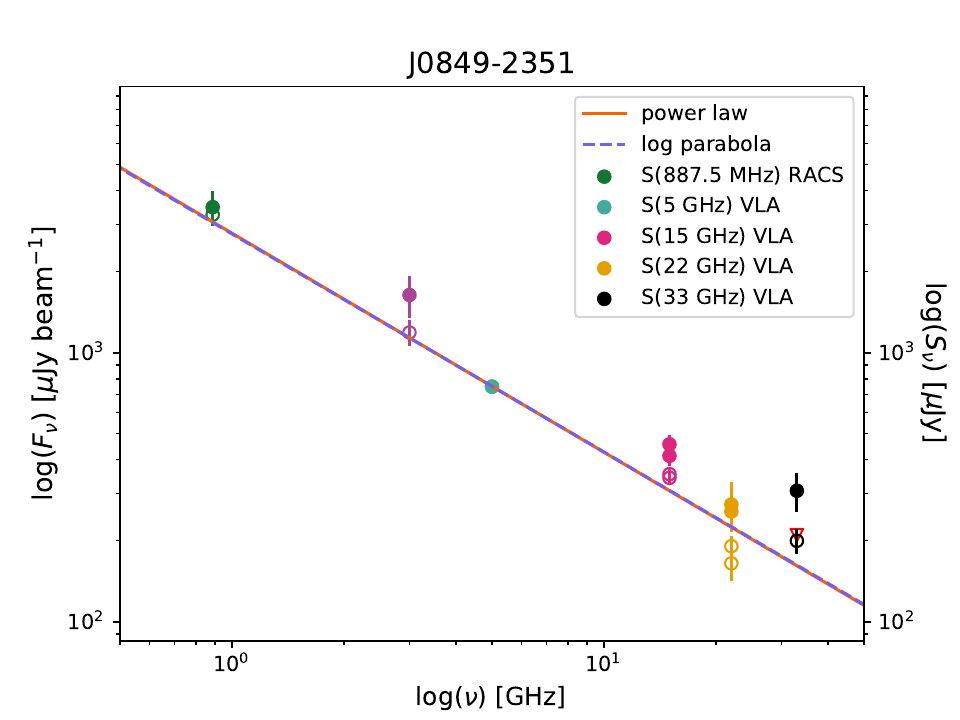}
    \caption{Spectrum of J0849-2351. Symbols as described in the text.}
    \label{fig:J0849-2351}
\end{figure}

\begin{figure}
    \centering
    \includegraphics[width=\columnwidth]{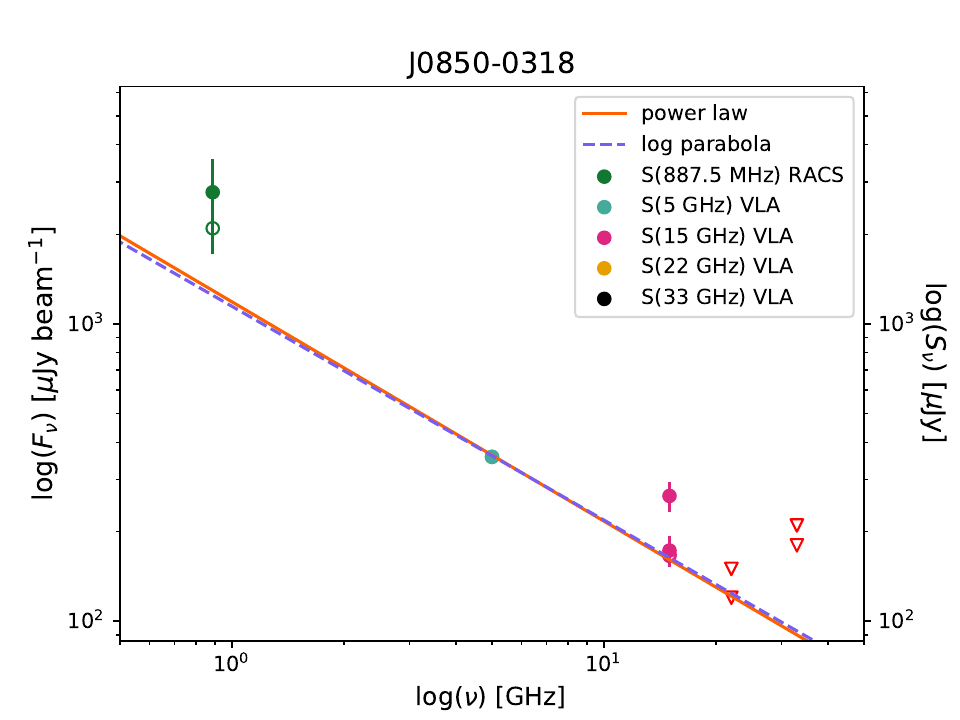}
    \caption{Spectrum of J0850-0318. Symbols as described in the text.}
    \label{fig:J0850-0318}
\end{figure}

\begin{figure}
    \centering
    \includegraphics[width=\columnwidth]{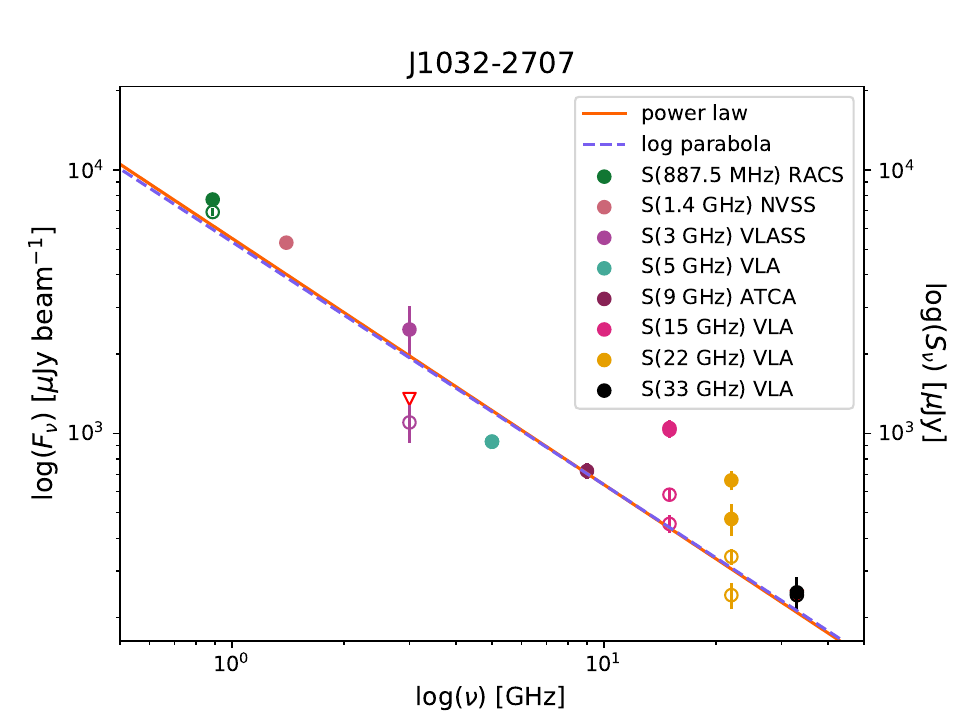}
    \caption{Spectrum of J1032-2707. Symbols as described in the text.}
    \label{fig:J1032-2707}
\end{figure}

\begin{figure}
    \centering
    \includegraphics[width=\columnwidth]{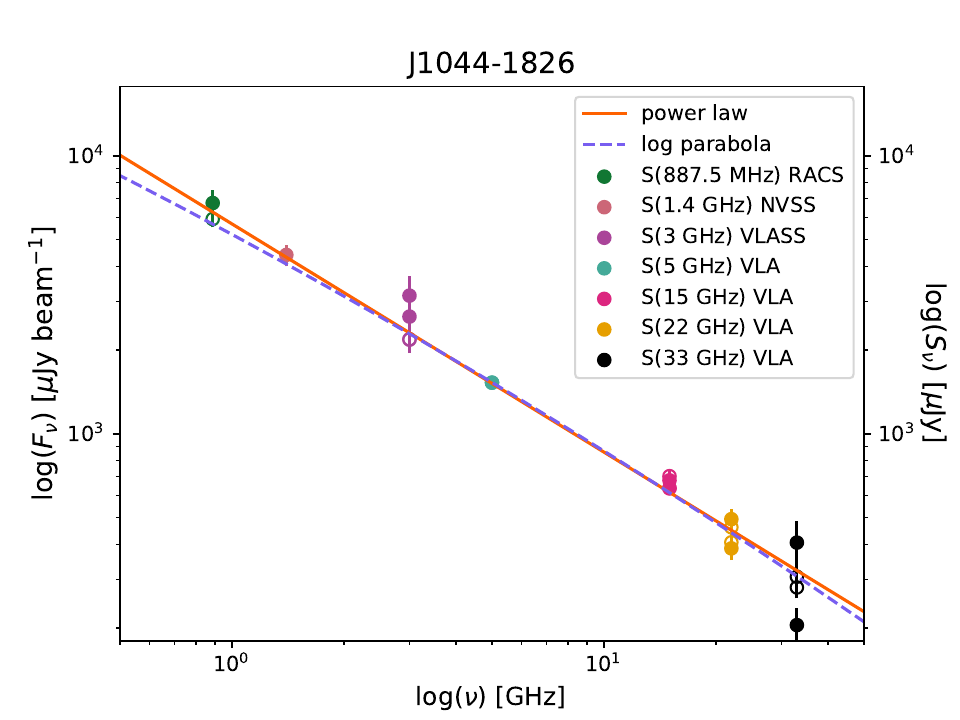}
    \caption{Spectrum of J1044-1826. Symbols as described in the text.}
    \label{fig:J1044-1826}
\end{figure}

\begin{figure}
    \centering
    \includegraphics[width=\columnwidth]{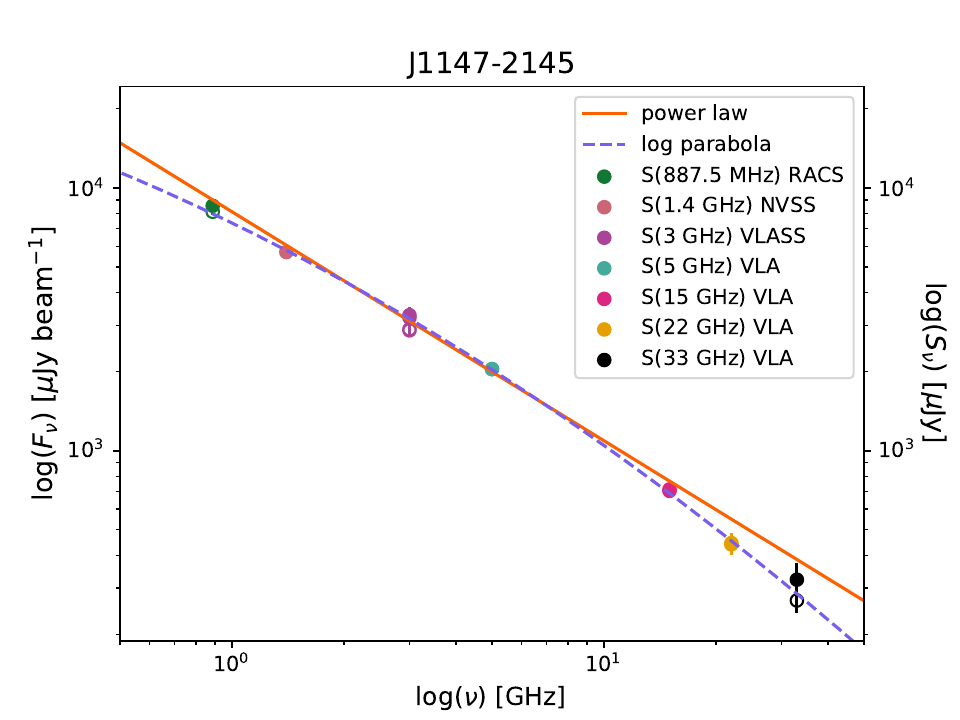}
    \caption{Spectrum of J1147-2145. Symbols as described in the text.}
    \label{fig:J1147-2145}
\end{figure}

\begin{figure}
    \centering
    \includegraphics[width=\columnwidth]{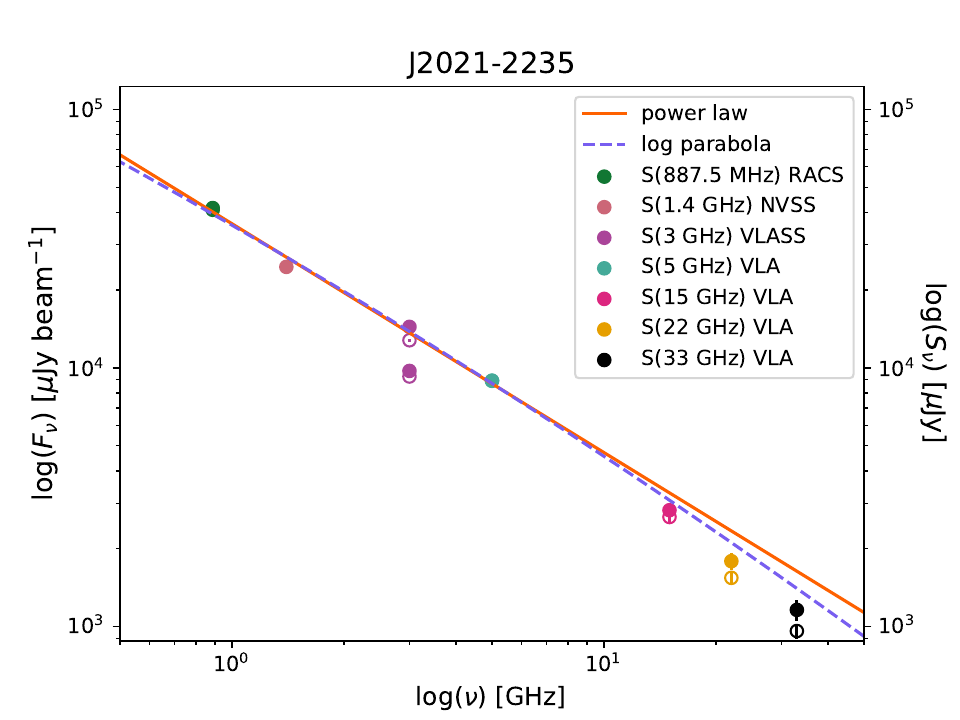}
    \caption{Spectrum of J2021-2235. Symbols as described in the text.}
    \label{fig:J2021-2235}
\end{figure}

\begin{figure}
    \centering
    \includegraphics[width=\columnwidth]{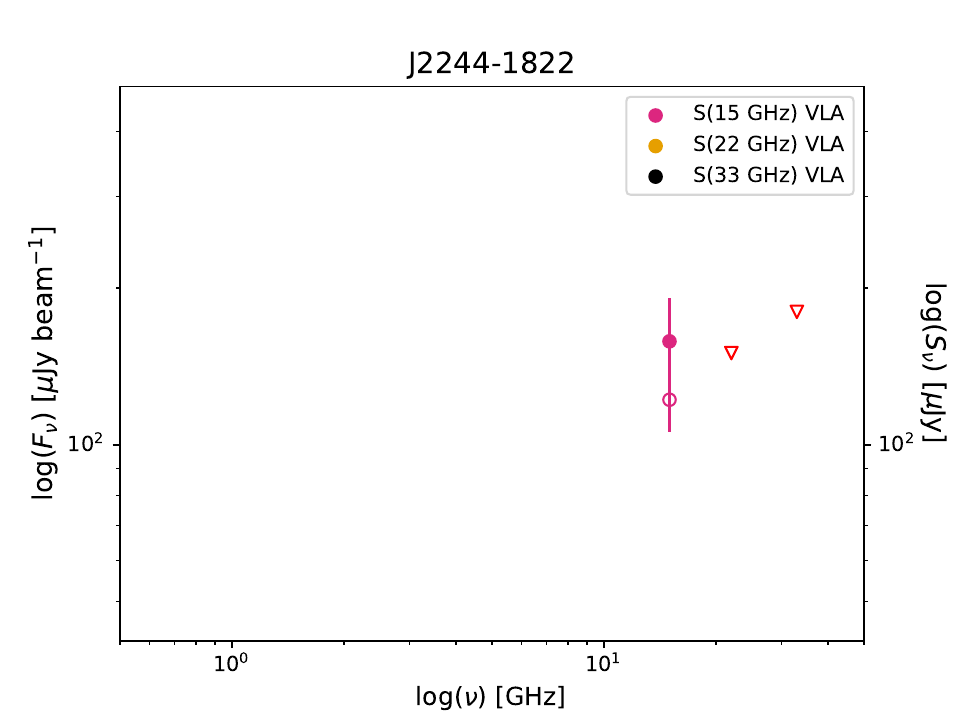}
    \caption{Spectrum of J2244-1822. Symbols as described in the text.}
    \label{fig:J2244-1822}
\end{figure}

\begin{figure}
    \centering
    \includegraphics[width=\columnwidth]{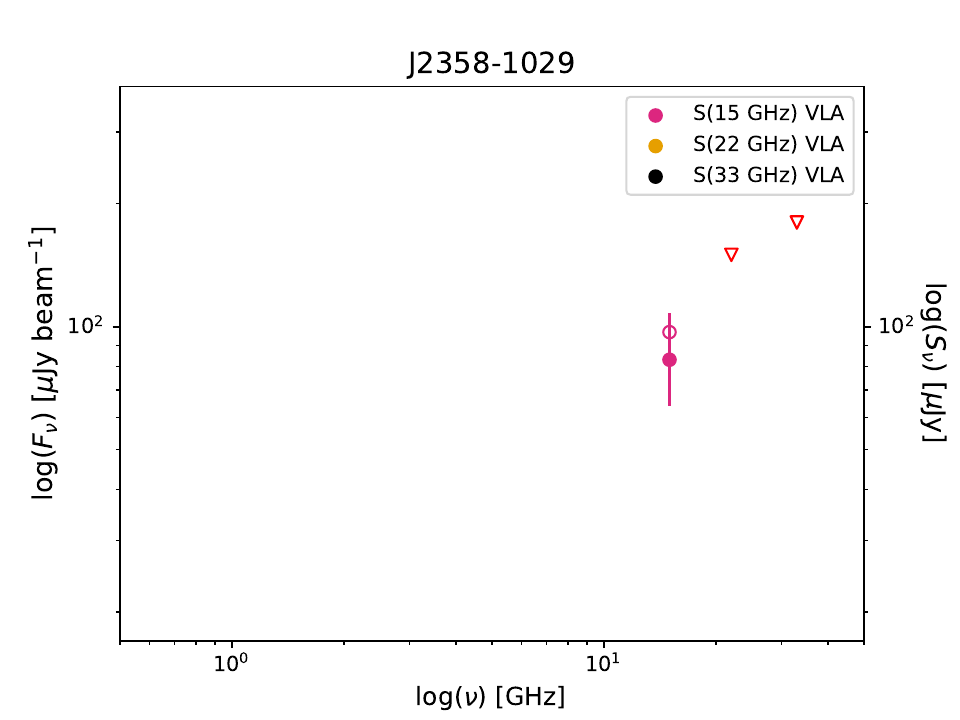}
    \caption{Spectrum of J2358-1029. Symbols as described in the text.}
    \label{fig:J2358-1029}
\end{figure}
%
\end{appendix}
\end{document}